\documentclass{aa}
\usepackage[varg]{txfonts}
\usepackage{graphicx}
\usepackage{natbib}
\usepackage{booktabs}
\usepackage{morefloats}
\bibpunct{(}{)}{;}{a}{}{,}

\begin{document}

\title{Monte-Carlo methods for NLTE spectral synthesis of supernovae.}

\author{M.~Ergon\inst{\ref{inst1}} \and C.~Fransson\inst{\ref{inst1}} \and A.~Jerkstrand\inst{\ref{inst4}} \and C.~Kozma\inst{\ref{inst1}} \and M.~Kromer\inst{\ref{inst2},\ref{inst3}} \and K.~Spricer\inst{\ref{inst5}}}

\institute{The Oskar Klein Centre, Department of Astronomy, AlbaNova, Stockholm University, 106 91 Stockholm, Sweden 
\label{inst1}
\and Max-Planck-Institut f{\"u}r Astrophysik, Karl-Schwarzschild-Str. 1, D-85748 Garching, Germany
\label{inst4}
\and Heidelberger Institut f{\"u}r Theoretische Studien, Schloss-Wolfsbrunnenweg 35, D-69118 Heidelberg, Germany
\label{inst2}
\and Institut f{\"u}r Theoretische Astrophysik am Zentrum f{\"u}r Astronomie der Universit{\"a}t Heidelberg, Philosophenweg 12, D-69120 Heidelberg, Germany
\label{inst3}
\and Department of Mathematics, Stockholm University, 106 91 Stockholm, Sweden
\label{inst5}}

\abstract{ We present JEKYLL, a new code for modelling of supernova (SN) spectra and lightcurves based on Monte-Carlo (MC) techniques for the radiative transfer. The code assumes spherical symmetry, homologous expansion and steady state for the matter, but is otherwise capable of solving the time-dependent radiative transfer problem in non-local-thermodynamic-equilibrium (NLTE). The method used was introduced in a series of papers by Lucy, but the full time-dependent NLTE capabilities of it have never been tested. Here, we have extended the method to include non-thermal excitation and ionization as well as charge-transfer and two-photon processes. Based on earlier work, the non-thermal rates are calculated by solving the Spencer-Fano equation. Using a method previously developed for the SUMO code, macroscopic mixing of the material is taken into account in a statistical sense. To save computational power a diffusion solver is used in the inner region, where the radiation field may be assumed to be thermalized. In addition, a statistical Markov-chain model is used to sample the emission frequency more efficiently, and we introduce a method to control the sampling of the radiation field, which is used to reduce the noise in the radiation field estimators. Except for a description of JEKYLL, we provide comparisons with the ARTIS, SUMO and CMFGEN codes, which show good agreement in the calculated spectra as well as the state of the gas. In particular, the comparison with CMFGEN, which is similar in terms of physics but uses a different technique, shows that the Lucy method does indeed converge in the time-dependent NLTE case. Finally, as an example of the time-dependent NLTE capabilities of JEKYLL, we present a model of a Type IIb SN, taken from a set of models presented and discussed in detail in an accompanying paper. Based on this model we investigate the effects of NLTE, in particular those arising from non-thermal excitation and ionization, and find strong effects even on the bolometric lightcurve. This highlights the need for full NLTE calculations when simulating the spectra and lightcurves of SNe. }


\titlerunning{Monte-Carlo methods for NLTE spectral synthesis of supenovae.}
\authorrunning{M. Ergon et al.}
\maketitle

\defcitealias{Maz93}{ML93}
\defcitealias{Luc02}{L02}
\defcitealias{Luc03}{L03}
\defcitealias{Luc05}{L05}
\defcitealias{Koz92}{KF92}
\defcitealias{Ker14}{K14}
\defcitealias{Jer15}{J15}
\defcitealias{Jer11}{J11}
\defcitealias{Jer12}{J12}
\defcitealias{Erg14}{E14}
\defcitealias{Erg15}{E15}
\defcitealias{Kro09}{K09}
\defcitealias{Sim07}{S07}
\defcitealias{Hil98}{H98}

\section{Introduction}
\label{s_intruduction}

Modelling the spectral evolution and lightcurves of supernovae (SNe) is crucial for our understanding of these phenomena, and much effort has been put into this during the last 50 years. To achieve realistic results local thermodynamic equilibrium (LTE) can generally not be assumed, and the full frequency-dependent, non-LTE (NLTE) problem has to be solved. Several paths exist and we have followed the one outlined in a series of papers by \citet[hereafter \citetalias{Luc02}, \citetalias{Luc03}, \citetalias{Luc05}]{Luc02,Luc03,Luc05}, in turn based on earlier work by \citet{Maz93} and \citet{Luc99}. Using this method, hereafter referred to as the Lucy method, the radiative transfer is solved by a Monte-Carlo (MC) calculation, which is alternated with a NLTE solution for the matter until convergence is achieved ($\Lambda$-iteration). Basic tests were performed in the original papers, and a simplified version of the method, assuming LTE for the population of excited states, has been implemented in the code ARTIS \citep[hereafter \citetalias{Kro09} and \citetalias{Sim07}]{Sim07,Kro09}. Several other codes, as for example TARDIS \citep{Ker14}, SEDONA \citep{Kas06}, SAMURAI \citep{Tan07} and the one by \citet{Maz00} are also based on the method, (or the early version of it), but are all restricted in one way or another. The Lucy method, or parts thereof, has also been used for modelling of other phenomena than SNe, for example in the code by \citet{Car06,Car08} and in the code by \citet{Lon02}, upgraded by \citet{Sim10}, both for radiative transfer in different kinds of circum-stellar disks.

Here, we present JEKYLL, a C++ based code which implements the full NLTE-version of the method, extended to include also non-thermal excitation and ionization as well as charge-transfer and two-photon processes. These extensions are particularly important for modelling in the nebular phase, and for the calculation of the non-thermal rates we have used the method developed by \citet[hereafter \citetalias{Koz92}]{Koz92}. Contrary to ARTIS, the initial version of JEKYLL is restricted to a spherical symmetric geometry. However, as the MC radiative transfer is performed in 3-D, this is not a fundamental restriction. The most fundamental limitation in JEKYLL, shared with most spectral codes, is the absence of hydrodynamics. As is often (but not always) justified, the ejecta is instead assumed to be in homologous expansion.

Another code based on $\Lambda$-iteration and MC radiative transfer is the steady-state NLTE code SUMO \citep[hereafter \citetalias{Jer11} and \citetalias{Jer12}]{Jer11,Jer12} aimed for the nebular phase. However, contrary to JEKYLL and other codes based on the Lucy method, conservation of MC packet energy is not enforced, so the MC technique used by SUMO is different. SUMO uses a novel statistical approach to represent the macroscopic mixing of the ejecta occurring in the explosion, which we have also adopted in JEKYLL.

In addition to the MC based codes, there is a group of codes that use finite difference techniques to solve the radiative transfer equation in a more traditional way. Examples of such codes are PHOENIX \citep{Hau99} and the general purpose NLTE code CMFGEN \citep[hereafter \citetalias{Hil98}]{Hil98}, which is similar to JEKYLL in terms of physics. In this paper, we compare JEKYLL with CMFGEN as well as with ARTIS and SUMO, which are also similar to JEKYLL in one way or another. These comparisons provide a thorough and critical test of the JEKYLL code. In a broader context, the comparison with CMFGEN also provides a test of the full time-dependent NLTE capabilities of the Lucy method.

In an accompanying paper (Ergon et al. in prep, hereafter Paper 2) we present an application of JEKYLL to Type IIb SNe, by modelling the early (before 150 days) evolution of a set of models previously evolved through the nebular phase with SUMO \citep[][hereafter \citetalias{Jer15}]{Jer15}. One of those is also presented in this paper as an example of a time-dependent NLTE calculation using a realistic ejecta model. However, for comparisons to observations and a deeper discussion of Type IIb SNe we refer to Paper 2.

The paper is organized as follows. In Sect.~\ref{s_physics} we discuss the underlying physical problem, and in Sect.~\ref{s_method} we describe the method used to solve this problem and the design of the code. In Sect.~\ref{s_comp} we provide the comparisons of JEKYLL with the ARTIS, SUMO and CMFGEN codes, and in Sect.~\ref{s_application} we provide the (example) application to Type IIb SNe, as well as some further tests based on it. Finally, in Sect.~\ref{s_conclusions} we conclude and summarize the paper.

\section{Physics}
\label{s_physics}

The general physical problem addressed is the time-evolution of the radiation field and the state of the matter, given the dynamical constraint of homologous expansion, and might be referred to as a radiation-thermodynamical problem. If the radiation field and the matter are in LTE this is simplified to a one-parameter (i.e. the temperature) problem, and may be easily solved. Otherwise, we are in the NLTE regime, and the number of parameters, as well as the complexity of the problem, increase drastically. 

As is often done, we solve for the radiation field and state of the matter separately, and the problem is split into a radiative transfer and a thermodynamical part. The coupling, provided by radiation-matter interactions, is enforced through $\Lambda$-iterations, where the state of the matter and the radiation field are alternately and iteratively determined from each other. The $\Lambda$-iteration concept is at the heart of the method, and in Sect.~\ref{s_lambda_iter} we provide some background and discuss the somewhat different meaning it has in traditional and MC based methods.

The state of the matter can be separated into a dynamical and thermodynamical part, where the former is trivially given by $\rho=\rho_{0}~(t/t_{0})^{-3}$ and $v=r/t$ through the constraint of homologous expansion. The thermodynamical part is given by the temperature, and the populations of ionized and excited states, which are solved for using the thermal energy equation and the NLTE rate equations, respectively. To simplify we assume steady state, which is motivated if the thermodynamical time-scale is much smaller than the dynamical time-scale. 

The radiation field is given by the specific intensity, which is solved for using an extended version of the MC based Lucy method, which is discussed in Sect.~\ref{s_mc_solver}. In a traditional code like CMFGEN, the specific intensity is solved for using the radiative transfer equation, whereas in a MC based code like JEKYLL, the radiative transfer is treated explicitly by propagating radiation packets which interact with the matter through absorption, emission and scattering. The different radiation-matter interactions supported are discussed in Sect.~\ref{s_radiation_matter}.

In addition, in SN ejecta radioactive decays emit high-energy photons or leptons, which give rise to a non-thermal electron distribution. Through collisions, these electrons contribute to the heating of the electron gas and the excitation and ionization of the ions. The problem may be broken up into two parts; deposition of the radioactive decay energy, and the partitioning of this energy into non-thermal heating, ionization and excitation. 

\subsection{$\Lambda$-iterations and convergence}
\label{s_lambda_iter}

In terms of the $\Lambda$-operator the radiative transfer equation may be written as $I=\Lambda[S]$\footnote{Historically the $\Lambda$-operator was defined in terms of the mean intensity, but this distinction does not matter for the discussion.}, where I is the intensity and S the source-function. If the source function depends on the intensity, as in the case of scattering, solving the problem requires inverting the $\Lambda$-operator. This is typically a costly operation, and we may instead try an iterative procedure called $\Lambda$-iteration \citep[see e.g.][]{Hub14}. In its original form an improved estimate of the intensity is then determined using the previous estimate of the source-function, that is $I_{i+1}=\Lambda[S_{i}]$. However, this method may converge extremely slowly if the source function is dominated by scattering, as the non-local coupling introduced only propagates one mean-free path per iteration. This may be solved by splitting the $\Lambda$-operator in two parts, one acting on the current iteration and one acting on the previous iteration, that is $I_{i+1}=\Lambda^{*}[S_{i+1}]+(\Lambda-\Lambda^{*})[S_{i}]$. With an appropriate choice of $\Lambda^{*}$, for example the local part of $\Lambda$, which is trivial to invert and still close to $\Lambda$, convergence could be accelerated, and the procedure is therefore known as accelerated $\Lambda$-iteration (see e.g.~\citealt{Can73a,Can73b}, \citealt{Sch84}, \citealt{Wer85} and~\citealt{Ols86}). 

It is important to realize that the explicit dependence of the scattering emissivity on the intensity does not cause slow convergence in the MC case. The reason for this is that the MC scattering emissivity depends directly on the current iteration of the MC radiation field. Actually, a MC $\Lambda$-iteration is similar to an accelerated $\Lambda$-iteration in the sense that the current iteration depends partly on itself. However, the implicit dependence of the MC emissivities on the intensity (via the matter quantities) may still cause slow convergence, and this is the problem addressed by the Lucy method. Enforcing the constraints of thermal and statistical equilibrium on the MC calculation, introduces a direct (but approximate) dependence of all MC emissivities on the current iteration of the MC radiation field. Although not formally proved, this is likely to accelerate the convergence in the general case, and as demonstrated in \citetalias{Luc03}, $\Lambda$-iterations based on this method have excellent convergence properties. 

We note, that most MC based methods use the Sobolev approximation, which helps to accelerate the convergence as line self absorption is already solved for. We also note, that in a steady-state calculation, all locations are causally connected to each-other, whereas in a time-dependent calculation the causally connected region grows with time, which makes convergence less demanding in each individual step of a time-dependent calculation than in a steady-state calculation.

\subsection{Statistical equilibrium}
\label{s_stat_eq}

To determine the populations of ionized and excited states, the NLTE rate equations need to be solved. Assuming steady state, these equations simplify to the equations of statistical equilibrium, where the rates of transitions in and out of each state are in equilibrium. The equation of statistical equilibrium for level $i$ of ion $I$ may be written
\begin{equation}
\label{eq_stat_equi}
\sum_{J=I \pm 1} r^{\text{BF}}_{J,j \rightarrow I,i}~n_{J,j}+\sum_{j \ne i} r^{\text{BB}}_{I,j \rightarrow i}~n_{I,j} = \left(\sum_{J=I \pm 1} r^{\text{BF}}_{I,i \rightarrow J,j} + \sum_{j \ne i} r^{\text{BB}}_{I,i \rightarrow j}\right)~n_{I,i}
\end{equation}
where $r$ is the rate (per particle) for bound-free (superscript $\text{BF}$) and bound-bound (superscript $\text{BB}$) transitions, and $n$ is the number density. We note, that the system of equations is non-linear as (some) transition rates (per particle) depend on the number densities. Transitions may be caused by absorption or emission of photons (Sect.~\ref{s_radiation_matter}), or by collisions involving ions and thermal (Sect.~\ref{s_collisions}) or non-thermal (Sect~\ref{s_rad_edep}) electrons.

\subsection{Thermal equilibrium}

To determine the thermal state of the gas the thermal energy equation needs to be solved. Assuming steady state, this equation simplifies to the equation of thermal equilibrium, where the heating and cooling of the gas are in equilibrium. The equation of thermal equilibrium may be written
\begin{equation}
\label{eq_therm_equi}
\sum_{J=I \pm 1} g^{\text{BF}}_{I,i \rightarrow J,j}(T)~n_{I,i} + \sum_{i \ne j} g^{\text{BB}}_{I,i \rightarrow j}(T)~n_{I,i} + \sum g^{\text{FF}}_{I}(T)~n_{I,i} = H^{\mathrm{NT}}
\end{equation}
where $g$ is the net heating rate (per particle) for bound-free (superscript $\text{BF}$), bound-bound (superscript $\text{BB}$) and free-free (superscript $\text{FF}$) transitions, and $H^{\mathrm{NT}}$ is the heating rate by non-thermal collisions. Heating/cooling may arise through absorption/emission of photons (Sect.~\ref{s_radiation_matter}), or through collisions involving ions and thermal (Sect.~\ref{s_collisions}) or non-thermal (Sect~\ref{s_rad_edep}) electrons.

\subsection{Radiation-matter interactions}
\label{s_radiation_matter}

In radiation-matter interactions, the radiation field and the matter (electrons and ions) exchange energy through absorption and emission of photons. Except for electron scattering, which is assumed to be coherent and isotropic in the co-moving frame (of the ejecta), and given by the Thomson cross-section, JEKYLL supports the following interactions.

\subsubsection{Bound-bound}

Through detailed balance, the excitation and de-excitation rates are related and determined by a single quantity, for example the spontaneous emission coefficient. We assume that the Sobolev approximation \citep{Sob57} applies, which is appropriate when expansion broadening dominates the thermal broadening. Expressions for the Sobolev optical depth as well as the transition rates are given in \citetalias{Luc02}. In addition, we also support de-excitation through two-photon emission for bound-bound transitions otherwise radiatively forbidden.

\subsubsection{Bound-free}

Through detailed balance, the ionization and recombination rates are related and determined by a single quantity, for example the photo-ionization cross-section. In bound-free transitions, the energy absorbed/emitted goes partly into ionization/recombination of the ion, and partly into heating/cooling of the electron gas. Expressions for the opacity, emissivity, transition rates and heating/cooling rates are given in \citetalias{Luc03}.

\subsubsection{Free-free (i.e.~bremsstrahlung)}

Assuming thermal matter, the opacity and emissivity are related through Kirchoffs law. 
In free-free interactions, the energy of the photons absorbed/emitted goes solely into heating/cooling of the electron gas. Expressions for the opacity, emissivity, and heating/cooling rates are given in \citetalias{Luc03}.

\subsection{Matter-matter interactions}
\label{s_collisions}

In matter-matter interactions, electrons and ions exchange energy through collisions. The collisions heat/cool the electron gas and result in bound-bound or bound-free transitions of the ions. Except for non-thermal collisions, which are discussed in Sect.~\ref{s_rad_epart}, JEKYLL supports the following interactions.

\subsubsection{Bound-bound and bound-free}

Through detailed balance, the collisional excitation and de-excitation rates are related and determined by a single quantity, for example the collisional strength. The same is true for the collisional ionization and recombination rates, and expressions for the transition rates and heating/cooling rates are given in \citetalias{Luc03}.

\subsubsection{Charge-transfer}

In collisions involving two ions, electrons may be transferred from one ion to another. This process is called charge-transfer and may be viewed as a recombination followed by a ionization. The charge transfer rates may be expressed in terms of a charge-transfer coefficient ($\alpha$) that depends only on the temperature as
\begin{equation}
\label{eq_col_trate_ct}
\begin{split}
&R_{\bar{I},\bar{J} \rightarrow \bar{U},\bar{L}}=\alpha_{\bar{I},\bar{J} \rightarrow \bar{U},\bar{L}}(T)~n_{\bar{I}}~n_{\bar{J}} \\
&R_{\bar{U},\bar{L} \rightarrow \bar{I},\bar{J}}={{\alpha_{\bar{I},\bar{J} \rightarrow \bar{U},\bar{L}}(T)} \over {\phi_{\bar{I},\bar{J},\bar{U},\bar{L}}(T)}}~n_{\bar{U}}~n_{\bar{L}}
\end{split}
\end{equation}
where $\phi_{\bar{I},\bar{J},\bar{U},\bar{L}}(T)=(n_{\bar{I}}^{\ast}~n_{\bar{J}}^{\ast})/(n_{\bar{U}}^{\ast}~n_{\bar{L}}^{\ast})$, the asterisk indicates the LTE value and $\bar{I}=(I,i)$ is an index vector specifying level $i$ of ion $I$. The energy difference between the initial and final state of the process gives rise to heating or cooling of the electron gas with a rate given by $R_{\bar{I},\bar{J} \rightarrow \bar{U},\bar{L}}~\left|\chi_{\bar{I},\bar{U}}-\chi_{\bar{L},\bar{J}}\right|$, where $\chi$ is the ionization energy.

\subsection{Radioactive decays}
\label{s_rad_decay}

\subsubsection{Energy deposition}
\label{s_rad_edep}

The energy released in the radioactive decays is carried by high-energy photons and leptons which deposit their energy in the ejecta mainly through Compton scattering on free and bound electrons. Although a detailed calculation is preferred, we use effective grey opacities determined through such calculations. We support the decay chains $\element[ ][56]{Ni}\! \rightarrow\! \element[ ][56]{Co}\! \rightarrow\! \element[ ][56]{Fe}$, $\element[ ][57]{Ni}\! \rightarrow\! \element[ ][57]{Co}\! \rightarrow\! \element[ ][57]{Fe}$ and $\element[ ][44]{Ti}\! \rightarrow\! \element[ ][44]{Sc}\! \rightarrow\! \element[ ][44]{Ca}$, which are the most important for core-collapse SNe. For these decays we adopt the life-times and energies from \citet{Koz98a} and the effective grey $\gamma$-ray opacities from \citetalias{Jer11}, and assume that the positrons emitted are locally absorbed.

\subsubsection{Energy partition}
\label{s_rad_epart}

Through a cascade of collisions the deposited energy gives rise to a high-energy tail on the otherwise Maxwellian electron distribution. The shape of the non-thermal electron distribution and the fractions of the energy going into heating, excitation and ionization through non-thermal collisions can be calculated by solving the Spencer-Fano equation (Boltzman equation for electrons). This problem was solved by \citetalias{Koz92} and for a further discussion we refer to this paper.

\section{Method and design}
\label{s_method}

Given the physical problem, we now describe the methods used to solve it, and provide an outline of how the code is designed. Except for the non-thermal solver, the code is written in C++, and the description therefore tends to reflect the object orientated structure of the code. The code is parallelized on a hybrid process (MPI) and thread (openMP) level, and we discuss this issue as well as the computational resources required in Sect.~\ref{s_resources}.

The SN ejecta are represented by a spatial grid of cells holding the local state of the matter and the radiation field. Although mostly geometry independent, the current version only supports spherically symmetric cells. To determine the state of the matter, JEKYLL provides several solvers with different levels of approximation (e.g.~LTE and NLTE), and to determine the radiation field it provides a MC solver based on the Lucy method. As discussed, through $\Lambda$-iterations the matter and the radiation field are alternately determined from each other, a procedure which in JEKYLL is terminated after a fixed but configurable number of iterations. JEKYLL also provides a diffusion solver, intended for use at high optical depths where the matter and radiation field may be assumed to be in LTE. JEKYLL may be configured to run in steady-state or time-dependent mode, although the latter only applies to the radiative transfer. Steady-state breaks down if the diffusion time is large, and is therefore best suited for modelling in the nebular (optically thin) phase, or of the SN atmosphere in the photospheric (optically thick) phase.

\subsection{Grid}
\label{s_grid}

The grid represents the SN ejecta and is spatially divided into a number of cells, which in the current version of the code are spherically symmetric. As mentioned, the code is mostly geometry independent, so cells with other geometries may easily be added in future versions. If macroscopic mixing is used, the cells may be further divided into compositional zones, geometrically realized as virtual cells. The grid provides functions to load the ejecta model, to load and save the state, as well as to export a broad range of derived quantities (e.g.~opacities).

\subsubsection{Cells}
\label{s_cell}

The cells hold the local state of the matter and the radiation field, and provide functions for the solvers to calculate derived quantities like opacities/emissivities and transition rates based on the local state and the atomic data. The local state of the matter is represented by the density, the temperature, and the number fractions of ionized and excited states. The local state of the radiation field is represented by the specific intensity, which is updated by the MC radiative transfer solver based on packet statistics following the method outlined by \citetalias{Luc03}. In addition, JEKYLL supports simplified radiation field models based on pure or diluted black-body radiation, given by $B_{\nu}(T_\mathrm{J})$ and $W B_{\nu}(T_\mathrm{R})$, respectively (see \citetalias{Kro09} for details). 

JEKYLL also allows the radiation-field to be approximated by the source-function as $I=S(\mathbf{n},T)$, which depends only on the local state of the matter. As discussed by \citet{Avr88}, this approximation is essentially a generalized version of the classical on-the-spot approximation. The generalized on-the-spot approximation is intended for use bluewards ground-state ionization edges, which typically have high optical depths and dominate the source-function. When using this approximation, only the bound-free opacities and emissivities are included in the source function, which is likely a good approximation.

\subsubsection{Virtual cells}
\label{s_virtual_cells}

JEKYLL implements the concept of virtual cells, introduced by \citetalias{Jer11} to account for macroscopic mixing on a grid otherwise spherically symmetric. Each cell may be divided into zones occupying some fraction (filling factor) of the cell volume, and otherwise geometrically unspecified. These zones may have different densities and compositions, and the state is solved for separately by the matter-state solver. With respect to the MC-solver the zones are represented by virtual cells differing only in a geometrical and statistical sense. The virtual cells are spherical, have a size corresponding to some number of clumps, and their location is randomly drawn during the MC radiative transfer based on their size and the zone filling factor.

\subsection{Atomic data}

Once converted to the JEKYLL format, any set of atomic data may be loaded from file. The data is organized in a hierarchical structure of atoms, their isotopes and ions, and the bound states of the ions. Each ion holds a list of bound-bound transitions, and each atom holds a list of bound-free transitions. The atomic data also contains an (optional) list of charge-transfer reactions, which are mapped on two bound-free transitions, one recombination and one ionization. The specific atomic data used for the comparisons in Sects.~\ref{s_comp_artis}-\ref{s_comp_cmfgen} are discussed in Appendix~\ref{a_conf_data_comp_artis}-\ref{a_conf_data_comp_cmfgen}. The default choice, used for the application in Sect.~\ref{s_application}, is inherited from SUMO (see \citetalias{Jer11} and \citetalias{Jer12}), and has been extended as described in Appendix \ref{a_conf_data_app}.

\subsection{MC radiative transfer solver}
\label{s_mc_solver}

The MC radiative transfer solver determines the radiation field, and is based on the Lucy method. The radiation field is discretized as packets (Sect.~\ref{s_packets}), which are propagated on the grid (Sect.~\ref{s_packet_propagation}) and interact with the matter (Sect.~\ref{s_packet_interaction}). We note, that the packets are propagated in 3-D, so the constraint of spherical symmetry only applies to the grid they are propagated on. In the calculation, the constraints of statistical and thermal equilibrium are enforced, which accelerates the convergence of the $\Lambda$-iterations (see Sect.~\ref{s_lambda_iter}). The original method has been extended to include non-thermal ionizations and excitations, as well as charge-transfer and two-photon processes. In addition, we introduce an alternative, more efficient way to draw the emission frequency (Sect.~\ref{s_markov_chains}), and a method to control the sampling of the radiation field (Sect.~\ref{s_packet_control}). Although we explain the basics, we refer to \citetalias{Luc02}-\citetalias{Luc05} for the details of the original method.

\subsubsection{Packets}
\label{s_packets}

The radiation field is discretized as packets, defined by their energy, frequency, position and direction. Following \citetalias{Luc03} and \citetalias{Kro09}, we classify these as r-, i-, k- and $\gamma$-packets. The packets are indivisible and indestructible (but see Sect.~\ref{s_packet_control} for a modified requirement), which enforce the constraint of thermal equilibrium on the MC calculation. Freely propagating photons are represented by r-packets, and upon absorption they are converted into i- and k-packets, representing ionization/excitation and thermal energy, respectively. The $\gamma$-packets are similar to the r-packets, but represent the $\gamma$-rays (or leptons) emitted in the radioactive decays, which are treated separately. Eventually, the i- and k-packets are converted into r-packets and re-emitted.

New r-packets are injected into the MC-calculation by sampling of the flux at the inner border (if any), and new $\gamma$-packets by sampling of the $\gamma$-ray emissivity. In addition, r-packets may be sampled from the initial intensity in each cell, as well as from the intensity in new cells taken over from the diffusion solver when the inner border is moved inwards.

\subsubsection{Propagation}
\label{s_packet_propagation}

When the r- and $\gamma$-packets are propagated they undergo physical (radiation-matter interactions) and geometrical (border crossings) events. Whereas propagation is carried out in the rest frame, the physical events take place in the co-moving frame, and the packets are transformed back and forth to $O(v/c)$. After each event, a random optical depth for the next physical event is drawn as $\tau=-~ln~z$, and the packet is propagated until the accumulated optical depth exceeds this value or a geometrical event occurs. We note, that line-absorption may only occur at the resonance distance, and the (Sobolev) line-opacity may be regarded as a delta-function. In the case of a physical event, the packet is processed as described in Sect.~\ref{s_packet_interaction}, and in either case propagation continues as described above. In the case of $\gamma$-packets effective grey opacities (Sect~\ref{s_rad_epart}) are used, which differs from the more detailed procedure by \citetalias{Luc05}. The r- and $\gamma$-packets leave the MC calculation by escaping through the outer border, where the r-packets are binned and summed to build the observed spectrum. When doing this light-travel time is taken into account by defining the observers time as $t_\mathrm{O}=t-(R/c)~\mu$, where R is the radius of the grid and $\mu$ the cosine of the angle between the packet direction and the radius vector.

If the packet enters a cell with macroscopic mixing of the ejecta (Sect.~\ref{s_virtual_cells}), a randomly orientated virtual cell is drawn based on the filling factors for the compositional zones (see \citetalias{Jer11} for details). As long as the packet remains in the cell, the distance to the next geometrical event is given by the size and the orientation of the virtual cell, and at each (virtual) border crossing the procedure is repeated.

\subsubsection{Interactions}
\label{s_packet_interaction}

Once the packet has been absorbed, an interaction process is drawn in proportion to the opacities. As mentioned, the interactions take place in the co-moving frame, and in the case of (coherent) electron scattering, the frequency does not change. Otherwise, an emission frequency is drawn using the method described by \citetalias{Luc02} and \citetalias{Luc03}, which enforces the constraints of statistical and thermal equilibrium on the MC calculation. Below we provide a summary of the original method and describe the extensions made for non-thermal, charge-transfer and two-photon processes. Before re-emission of the packet a new direction is drawn from an isotropic distribution.

\begin{figure}[tbp!]
\includegraphics[width=0.5\textwidth,angle=0]{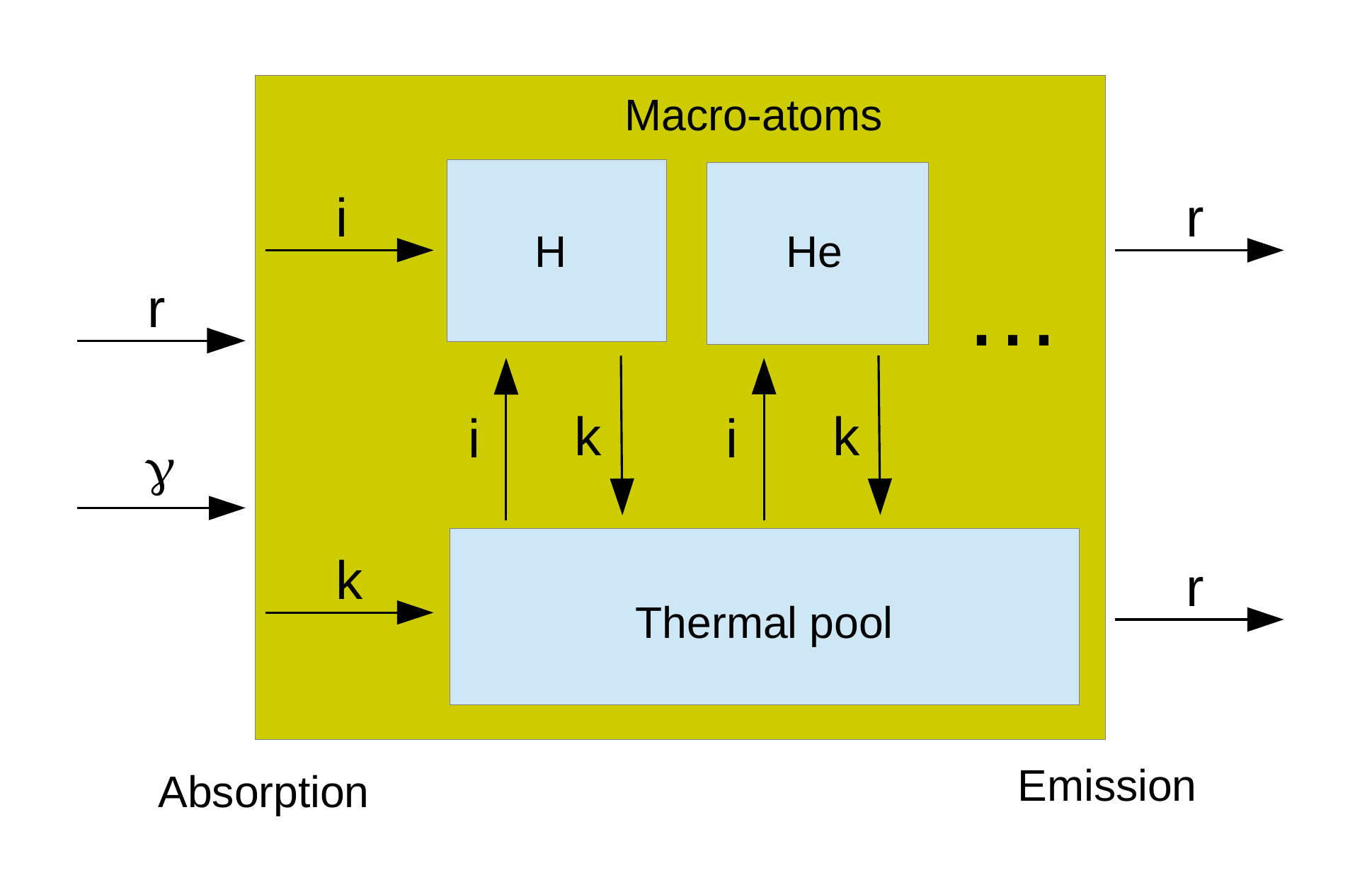}
\caption{Schematic figure of the MC state machine, showing the macro-atoms and the thermal pool, as well as the flows of r-, $\gamma$-, i- and k-packets.}
\label{f_mc_state_machine}
\end{figure}

\paragraph{Original method}

To enforce the aforementioned constraints on the MC calculation, \citetalias{Luc02} and \citetalias{Luc03} introduce the concepts of macro-atoms and the thermal pool, which are the MC analogues of the equations of statistical and thermal equilibrium. In computational terminology, a macro-atom is a (finite) state-machine, consisting of a set of states and the rules that govern internal transitions (between the states) and de-activation (of the machine). The states of a macro-atom corresponds to the energy levels of an atomic specie, and it is activated by ionizations and excitations and de-activated by recombinations and de-excitations. The rules that govern internal transitions and de-activation of a macro-atom are based on the equations of statistical equilibrium, and we give the details below. Together, the macro-atoms and the thermal pool constitute a single hierarchical state-machine, which we refer to as the MC state-machine. In this context, the macro-atoms are sub-machines, and may be regarded as states with respect to a base-machine. In the base-machine, collisional activation and de-activation of the macro-atom sub-machines correspond to internal transitions between the thermal pool and the macro-atom states, and the rules that govern transitions in and out of the thermal pool are based on the equation of thermal equilibrium. The MC state-machine is activated by an absorption of an r- or $\gamma$-packet, and de-activated by the emission of an r-packet, and is illustrated in Fig.~\ref{f_mc_state_machine}, where we also indicate the flows of i-, and k-packets within the machine.

Following the packets through the MC state-machine, the absorbed r- and $\gamma$-packets are converted into k- or i-packets in proportion to the energy going into heating and ionization/excitation\footnote{Only the heating-channel is allowed for $\gamma$-packets in the original method.}. In the former case, the k-packets are transferred to the thermal pool, and in the latter case, the i-packets activate the macro-atoms through radiative ionizations and excitations, drawn in proportion to their opacities. Eventually, the macro-atoms de-activates, and in de-activations through radiative transitions, the i-packets are converted to r-packets and re-emitted, whereas in de-activations through collisional transitions, the i-packets are converted to k-packets and transferred to the thermal pool. The k-packets enter the thermal pool through radiative and collisional heating and leave through radiative and collisional cooling, in which case they are converted into r- or i-packets in proportion to the cooling rates. In the former case, the emission processes are drawn in proportion to their cooling rates and the r-packets are re-emitted, and in the latter case the i-packets activate the macro-atoms through collisional ionizations and excitations, drawn in proportion to their cooling rates. We note, that before the r-packets are re-emitted, their frequencies are drawn from the (normalized) emissivities of the de-activating processes.

Although the method is conceptually simple, it is a bit involved in the details, in particular with respect to the macro-atoms. As described in \citetalias{Luc02}, the rules for the macro-atom state-machine are derived through a rewrite of the equations of statistical equilibrium in terms of energy. This leads to a number of terms that can be identified as the energy rates for activations, internal transitions and de-activations, and from this the probabilities can be calculated. A macro-atom is activated in level $i$ by an physical upward transition (e.g.~excitation) to this this level. Once in level $i$, each physical transition with number rate $R_{i \rightarrow j}$ corresponds to an internal state-machine transition with probability $P^{\mathrm{I}}_{i \rightarrow j} \propto R_{i \rightarrow j}~E_{l}$ (\citetalias{Luc02}:~Eq.~9), where $E_{l}$ is the energy of level $l=min(i,j)$. In addition, each physical downward transition (e.g. de-excitation) may de-activate the macro-atom with probability $P^{\mathrm{D}}_{i \rightarrow l} \propto R_{i \rightarrow l}~(E_{i}-E_{l})$ (\citetalias{Luc02}:~Eq.~7). If an internal transition is drawn, the state-machine proceeds to level $j$ and the procedure is repeated. We note, that internal and de-activating transitions may result from several physical processes, and if not otherwise stated, $P^{\mathrm{I}}_{i \rightarrow j}$ and $P^{\mathrm{D}}_{i \rightarrow l}$ refer to the total probabilities for such transitions.

\paragraph{Non-thermal processes}

Upon absorption, $\gamma$-packets are converted into k- and i-packets in proportion to the energy going into heating and ionization/excitation. In the former case, the k-pakets are transferred to the thermal pool, and in the latter case, the i-packets activate the macro-atom state-machines by non-thermal transitions drawn in proportion to their energy rates. In the original method, only the heating-channel was allowed, and the addition of the ionization and excitation channels is one of our most important extensions to the Lucy method. The macro-atom state-machines are modified by adding non-thermal transitions, where the probabilities are calculated from their number rates as explained above. Non-thermal transitions are upward, and therefore correspond to internal transitions.

\paragraph{Charge-transfer processes}

As mentioned, charge-transfer is a collisional process that may be viewed as a recombination followed by an ionization, where the (small) energy difference results in either heating or cooling. The macro-atom state-machines are therefore modified by adding the corresponding ionizations and recombinations, where the probabilities are calculated from their number rates as explained above. Charge-transfer ionizations correspond to internal transitions, whereas charge-transfer recombinations correspond to internal and de-activating transitions. 

De-activation of a macro-atom state-machine through a charge-transfer recombination results in either activation of another macro-atom state-machine through the corresponding ionization or in the conversion of the i-packet into a k-packet. The latter corresponds to the conversion of ionization energy into thermal energy, which may only happen if the reaction is exo-thermic, and is drawn in proportion to the energy going into heating. Correspondingly, if the reaction is endo-thermic, k-packets may be converted into i-packets, in which case a macro-atom state-machine is activated by the corresponding ionization. This corresponds to the conversion of thermal energy into ionization energy, and is drawn in proportion to the cooling rate as described above.

\paragraph{Two-photon processes}

The macro-atom state-machines are modified by adding two-photon transitions, where the probabilities are calculated from their number rates as explained above. Two-photon transitions are downward, and might therefore be either internal or de-activating, and in the latter case the emission frequency is drawn from the (normalized) two-photon emissivity.

\subsubsection{Markov-chain solution to the MC state-machine}
\label{s_markov_chains}

A problem with the original method is that the number of transitions in the MC state-machine may become very large. This is particularly true when the collisional rates are high, causing the state-machine to bounce back and forth between macro-atoms and the thermal pool. To avoid this we introduce a statistical Markov-chain model to calculate the probability that the state-machine de-activates from a given state. This allows the machine to proceed to the state from which it de-activates in a single draw. A Markov-chain model can be constructed for the complete MC state-machine, as well as for its base- and sub-machine parts. Markov chain statistics have been used for MC radiative transfer before, for example for scattering in planetary atmospheres by \citet{Esp78}, but the application of it to the Lucy method described here is novel.

In the case of a macro-atom sub-machine, the state-machine consists of $N$ states corresponding to the energy levels of an atomic specie, where the probabilities for internal and de-activating transitions, $P^{\mathrm{I}}_{i \rightarrow j}$ and $P^{\mathrm{D}}_{i \rightarrow j}$, are calculated as described in Sect.~\ref{s_packet_interaction}. In Markov-chain terminology, states are called transient if they are visited a finite number of times, recurrent if they are visited an infinite number of times, and absorbing if they can not be left. As the macro-atom will eventually de-activate, each state is only visited a finite number of times, and corresponds to a transient state in our Markov-chain model. In addition, we associate each transient state with an absorbing state, where a transition from a transient state into its associated absorbing state represents de-activation of the macro-atom (through any of the de-activating transitions).

The probabilities for transitions between transient states (i.e.~internal transitions in the Lucy terminology) form the matrix $\mathbf{P}^{\mathrm{T}}$, where $P^{\mathrm{T}}_{i,j}=P^{\mathrm{I}}_{i \rightarrow j}$ is the probability for a transition from the transient state $i$ into the transient state $j$. The probabilities for transitions from transient states into their associated absorbing states (i.e.~de-activations in the Lucy terminology) forms the diagonal matrix $\mathbf{R}$, where $R_{i,i}=\sum P^{\mathrm{D}}_{i \rightarrow j}$ is the probability for a transition from the transient state $i$ into its associated absorbing state $i$. From Markov-chain theory \citep[see e.g.][]{Ros07} we obtain the probability to end up in the absorbing state $j$ given that the macro-atom is activated in the transient state $i$ as $(\mathbf{SR})_{i,j}$, where the matrix $\mathbf{S}$ is given by
\begin{equation}
\mathbf{S}=(\mathbf{I}-\mathbf{P}^{\mathrm{T}})^{-1}
\end{equation}
where $\mathbf{I}$ is the identity matrix. Therefore, to find the probability that the macro-atom de-activates from state $j$ given that it was activated in state $i$, it is only neccesary to look up the $j$:th element in the $i$:th row of the $\mathbf{SR}$ matrix. 
Once the state from which the macro-atom de-activates has been drawn, the de-activating transition is drawn from their (normalized) probabilities. The implementation is based on two look-up tables for each state, one containing (a row of) $\mathbf{SR}$, and one containing the (normalized) de-activating transition probabilities. Once a macro-atom is activated, the state from which it de-activates is drawn from the former table, the state-machine proceeds to this state, and the de-activating transition is drawn from the latter table.

The Markov-chain model for a macro-atom sub-machine described above is easily generalized to the complete MC state-machine, in which case the states are the energy levels of all macro-atoms plus the thermal pool. We note, however, that collisional activation and de-activation of the macro-atoms then correspond to internal transitions between the thermal pool and the macro-atoms. In case the method is applied to the complete MC state-machine, the $\mathbf{S}$ matrix has size $N\times N$, where $N$ is the total number of energy levels for all macro-atoms plus one (the thermal pool), and the computational time to invert the matrix is a potential problem. This may be circumvented by splitting the MC state-machine into its base- and sub-machine parts, and calculate the corresponding $\mathbf{S}$ matrices separately. The procedure is similar to what is described above, but the computational time to invert the base- and sub-machine $\mathbf{S}$ matrices is much less than for the complete $\mathbf{S}$ matrix. We give the details on the split state-machine approach in Appendix~\ref{a_markov_chains}.

\subsubsection{Packet sampling control}
\label{s_packet_control}

\begin{figure}[tbp!]
\includegraphics[width=0.5\textwidth,angle=0]{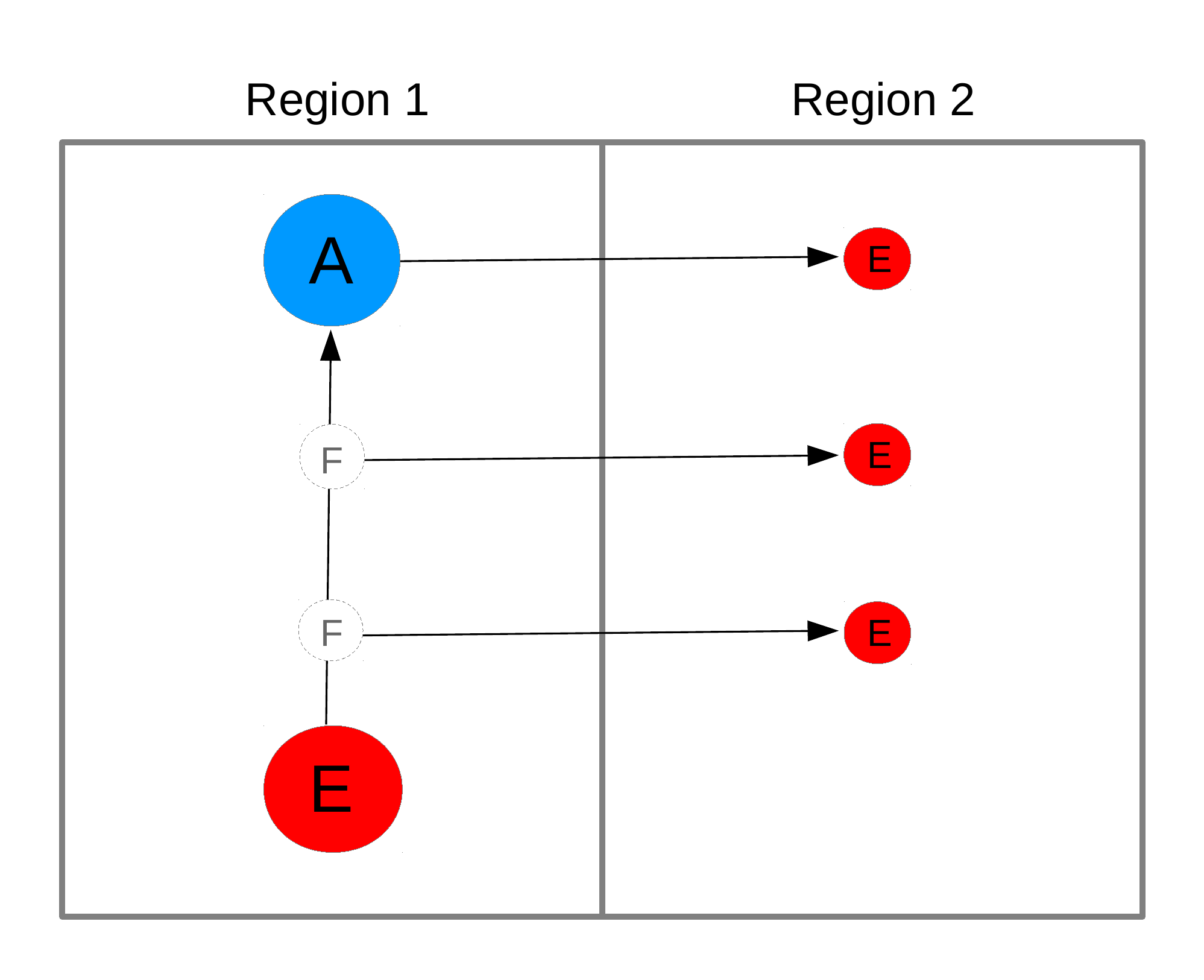}
\caption{Schematic figure of the re-sampling procedure for physical events. A packet emitted in region 1 with a large packet size (left), give rise to emission in region 2 with a small packet size (right) through fictitious absorption, and is eventually physically absorbed. Regions 1 and 2 are assumed to be spatially coincident but separated in frequency. The circles indicates the size of the packets, emission is labelled with an E, and physical and fictitious absorption is labelled with an A and an F, respectively.}
\label{f_packet_control}
\end{figure}

Another problem with the original method is that there is no (or limited) control of the number of packets as a function of frequency, space and time. This may result in too few packets, leading to noise in the radiation field estimators, or too many, leading to unnecessary computational effort. The number of packets can not be directly controlled, but this might instead be achieved by adjusting the amount of energy they carry. Hereafter, we refer to this as their size, and by  conservation of energy the number of packets is inversely proportional to their size. We therefore introduce a method for continuous re-sampling of the radiation field through control of the packet size, which is allowed to vary as a function of frequency, space and time. This breaks the indivisibility and destructibility requirements introduced by \citetalias{Luc02}, but conservation of energy, which is the essential physical property, is still maintained in an average sense.

A set of sampling regions (bounded in frequency, space and time) is defined, and each of these is assigned a packet size. When packets flow from one sampling region into another, their size is adjusted to that of the destination region. To maintain the rate of energy flowing into the destination region, the rate of packets flowing into it has to be adjusted with $F$, the ratio of the packet sizes in the source and destination regions. Consider now the series of events that occurs. In the original method, a leave event in the source region triggers an enter event in the destination region. However, as the number of leave and enter events in our method may differ, this procedure no longer apply. The simplest method to solve this, which is used for geometrical events, is to trigger $F$ enter events for each leave event. More specifically, when crossing a border, the packets are split into $F$ child packets\footnote{Ignoring here the fact that $F$ is not an integer. In practice, the number of child packets is drawn to fulfil this condition in an average sense.} (if $F$>1) or terminated with probability 1-$F$ (if $F$<1).

In the case of physical events, when the frequency changes, packets leave the source region through absorption and enter the destination region through emission, and the emission rates need to be adjusted with $F$. A fundamental short-coming of the method used for geometrical events is that the number of events in the source region is not adjusted, and this number may be small (or even worse, it may be zero). To overcome this, we introduce a class of fictitious absorption events, which occur $F$ times as often as the physical absorption events, and which have the sole purpose to trigger the emission events. For an interaction process with physical opacity $\kappa$, we may then define a fictitious opacity $\kappa_{\mathrm{F}}$  corresponding to these fictitious absorption events. The interaction rate given by $\kappa_{\mathrm{F}}$ may be higher or lower than the one given by $\kappa$, and is the one required to produce the adjusted emission rates\footnote{Strictly speaking, this is only true as long as physical absorption proceeds at the original rate.}.

As the emission rate is adjusted, whereas physical absorption needs to proceed at the original rate, the fictitious opacity to use is not $\kappa_{\mathrm{F}}$ but $max(\kappa_{\mathrm{F}},\kappa)$. Using this fictitious opacity, a packet is propagated as described in Sect.~\ref{s_packet_propagation}, and once an interaction is drawn, it is selected for absorption and emission with probabilities $max(\kappa/\kappa_{\mathrm{F}},1)$, and $max(\kappa_{\mathrm{F}}/\kappa,1)$,  respectively. If the packet is selected for absorption but not for emission it is terminated, and if the packet is selected for emission but not for absorption a child packet is created. Otherwise the interaction is handled as described in Sect.~\ref{s_packet_interaction} (which also recovers the normal behaviour if $\kappa=\kappa_{\mathrm{F}}$). As mentioned, the size of the emitted packets are always adjusted to that of the destination region. The re-sampling procedure for physical events outlined here is illustrated in Fig.~\ref{f_packet_control}, and as is possible to show, it gives the correct (average) energy flows in and out of each sampling region.

Although the basic idea is straightforward, the actual implementation is complicated by the way the emission frequency is drawn in the Lucy method (Sect.~\ref{s_packet_interaction}).  In particular, the probabilities for all possible paths a packet may take through the MC state-machine need to be adjusted. This is only possible if these are known, which is only the case if a Markov-chain solution to the MC state-machine (Sect.~\ref{s_markov_chains}) has been obtained. We give the details of the implementation in Appendix~\ref{a_packet_control}, where we explain how to adjust the MC state-machine probabilities, and how to calculate the fictitious opacity. 

When using the method  in JEKYLL, the number of packets is controlled by an adaptive algorithm, where the packet size in each sampling region is adjusted once per $\Lambda$-iteration. Assuming a Poisson distribution, the signal-to-noise ratio (SNR) in each sampling region is estimated based on the mean number of (physical and geometrical) packet events per frequency bin. Comparing this to a pre-configured target SNR, the packet size in each sampling region is adjusted based on the ratio of the target and estimated SNR. In addition, the packet size is bounded below and above by pre-configured minimum and maximum values.

\subsection{Matter state solvers}

To determine the state of the matter, JEKYLL provides the NLTE solver, as well as the more approximate LTE and \citet[][hereafter ML93]{Maz93} solvers. It also provides an option to mix these solvers, for example by using the NLTE solver for the ionization and the LTE solver for the excitation, in a manner similar to what is done in ARTIS. In addition, JEKYLL provides a solver to determine the non-thermal electron distribution, used by the NLTE solver.

\subsubsection{LTE solver}

The LTE solver determines the state of the matter assuming that LTE applies. The populations of ionized and excited states are calculated using the Saha ionization and Boltzman excitation equations, respectively. The temperature used may be that associated with the pure or diluted black-body radiation field models ($T_\mathrm{J}$ or $T_\mathrm{R}$; see Sect.~\ref{s_cell} and \citetalias{Kro09}), or the matter temperature determined by some other method (e.g.~thermal equilibrium).

\subsubsection{ML93 solver}

The ML93 solver determines the state of the matter assuming that the radiative rates dominate, and is based on the approximations for the populations of ionized and excited states derived by \citet{Maz93} and \citet{Abb85}. Following \citet{Maz93}, the temperature is assumed to be controlled by the radiation field and set to $0.9 T_\mathrm{R}$, where $T_\mathrm{R}$ is the temperature associated with the diluted black-body radiation field model (see Sect.~\ref{s_cell} and \citetalias{Kro09}).

\subsubsection{NLTE solver}

The NLTE solver determines the state of the matter by solving the equations of statistical and thermal equilibrium for the level populations and the temperature, respectively. The solution is determined in two steps. First, thermal equilibrium is scanned for in a configurable temperature interval (centered on the solution from the previous $\Lambda$-iteration)  using the bi-section method. In doing this, statistical equilibrium is solved for at each temperature step. Based on this estimate, thermal and statistical equilibrium are simultaneously iterated for until convergence is achieved, using a procedure similar to what is described by \citetalias{Luc03}.

\paragraph{Statistical equilibrium}

The non-linear system of statistical equilibrium equations (Eq.~\ref{eq_stat_equi}) is solved by iteration on the level populations. In each step the system is linearized in terms of changes in the level populations, and the rates and their derivatives are calculated using the previous estimate of these. The linearized system is then solved for changes in the level populations using lower-upper (LU) decomposition and back-substitution. If all number derivatives (explicit and implicit) are included this is equivalent to a Newton-Raphson solver, but in JEKYLL this is configurable, and in the simplest configuration only the explicit derivatives (i.e.~rates per particle) are included. 

The system of equations may be solved separately for the states of each atom, ignoring any coupling terms, or for all states at once. As the total number of states may be too large for a coupled solution, there is also a possibility to alternate a decoupled solution with a fully coupled solution for the ionization balance. Typically a decoupled solution works well, but charge-transfer reactions and the source-function radiation field model (see Sect.~\ref{s_cell}) may introduce strong coupling terms. Transition rates (Sects.~\ref{s_radiation_matter} and \ref{s_collisions}) for bound-bound and bound-free radiative and collisional processes, as well as for non-thermal, charge-transfer and two-photon processes are all supported, but which ones to include is a configurable choice. 

\paragraph{Thermal equilibrium}

The equation of thermal equilibrium (Eq.~\ref{eq_therm_equi}) is solved either using the bisection method (initial estimate) or Newton-Raphson's method (refined estimate), in which case an explicit temperature derivative is used. Heating and cooling rates (Sects.~\ref{s_radiation_matter} and \ref{s_collisions}) for bound-bound and bound-free radiative and collisional processes, free-free processes, as well as non-thermal and charge-transfer processes are all supported, but which ones to include is a configurable choice. 

\subsubsection{Non-thermal solver}

The non-thermal solver determines the non-thermal electron distribution resulting from the radioactive decays, and the fraction of the deposited energy going into heating, excitation and ionization. This is done by solving the Spencer-Fano equation (i.e. the Boltzman equation for electrons) as described in \citetalias{Koz92}. The fractions going into heating, excitation and ionization depend on the electron fraction, but as this dependence is rather weak a decoupled solution works well. The non-thermal solver is called by the NLTE solver at least twice each $\Lambda$-iteration (before each of its main steps), and otherwise whenever the electron fraction changes more than some pre-configured value.

\subsection{Diffusion solver}
\label{s_diffusion_solver}

The diffusion solver determines the temperature in each cell by solving the thermal energy equation assuming spherical symmetry, homologous expansion, LTE and the diffusion approximation for the radiative flux. This results in a non-linear system of equations for the temperature in each cell, which is solved by a Newton-Raphson like technique, similar to the one used by \citet{Fal77}. Two specific topics require some further discussion though; the Rosseland mean opacity used in the diffusion approximation, and the boundary where the diffusion solver is connected to the MC radiative transfer solver.

\subsubsection{Opacity}

The Rosseland mean opacity used in the diffusion approximation is calculated from the LTE state of the matter and the atomic data. This may sound straightforward, but the bound-bound opacity, and in particular the macroscopic mixing (see Sect~\ref{s_virtual_cells}) complicates things. In the latter case, if the clumps are all optically thin, the opacity may be calculated as an average over the compositional zones, but otherwise a geometrical aspect enters the problem. Therefore a Monte-Carlo method is used to calculate the Rosseland mean opacity. In each cell a large number of packets are sampled based on the black-body flux distribution and the filling factors for the compositional zones. These packets are then followed until they are absorbed, and their path-length averaged to get the Rosseland mean free path. This gives the Rosseland mean opacity, including the bound-bound contribution, as well as the geometrical effects arising in a clumpy material. 

\subsubsection{Connecting boundary}

If the diffusion solver is connected to the MC radiative transfer solver, appropriate boundary conditions must be specified for both solvers. As connecting boundary condition for the diffusion solver we have used the temperature in the innermost cell handled by the MC radiative transfer solver. As connecting boundary condition for the MC radiative transfer solver we have used the luminosity at this boundary determined with the diffusion solver. This is analogous to how the boundary between the diffusion and radiative transfer solvers is treated in \citet{Fal77}, except that in JEKYLL these calculations are not coupled and performed separately. To implement the connecting boundary condition for the MC radiative transfer solver an approximate method is used. During a time-step $\Delta t$, packets with total energy $L \Delta t$ are injected at the connecting boundary, whereas packets propagating inwards are simply reflected at this boundary. The frequency of the injected packets are sampled from a black-body distribution at the temperature of the innermost cell.

\subsection{Notes on the MC radiation field}

As mentioned in Sect,~\ref{s_packet_control}, the sampling of the MC radiation field is a potential problem with the original method. In our experience this problem is most severe bluewards $\sim$3000 $\AA$~and in the outer region of the ejecta. In principle, this could be solved by the method for packet sampling control, but in practice it is better to use the generalized on-the-spot approximation (Sect.~\ref{s_cell}) bluewards ground-state ionization edges of abundant species (e.g.~the Lyman break). The reason for this is twofold. First, to achieve a reasonable SNR in this region might require very large boost factors, which could potentially make the method for packet sampling control unstable. Second, in case a residual from two almost cancelling radiative rates (e.g.~ionization and recombination in the Lyman continuum) is large enough to be important for the solution, even larger boost factors might be needed to achieve the required SNR. Due to this we have used both packet sampling control and the generalized on-the-spot approximation in most of the simulations presented here and in Paper 2.

\subsection{Computional resources and parallelization}
\label{s_resources}

JEKYLL is solving a complex problem with several thousands of independent quantities, so the computational resources (e.g.~CPU time and memory) required to run a simulation are inevitably large. To handle this, the code is parallelized and uses several methods to reduce the computational effort. The code is parallelized on a hybrid process (MPI) and thread (openMP) level, although the number of threads are limited by shared memory access. The CPU time required for the MC radiative transfer is (roughly) proportional to the number of MC packets. As these are independent, the MC radiative transfer is naturally parallelized on them, and the processing power scales nicely with the number of CPU cores up to some large number of MC packets.

The CPU time required for the NLTE solver is (roughly) proportional to the number of grid cells, and as these are independent, the NLTE solver is naturally parallelized on them. However, this limits the scaling of the processing power to the number of cells (which is much smaller than the number of MC packets). Therefore the code is further parallelized, which increases the scaling limit with a factor of at least a few. We note, that the CPU time needed to solve the statistical equilibrium equations and to invert the Markov Chain $S$-matrix (see Sect.~\ref{s_markov_chains}) is proportional to the third power of the number of energy levels, so the number of such levels is critical.

The CPU time needed for a typical simulation like the Type IIb model presented in Sect.~\ref{s_application} is a few thousand CPU hours, which using a few hundred CPU cores results in an execution time of about ten hours. The number of MC packets used (in each iteration) in such a typical simulation is a few millions, the number of grid cells between 50 and 100 and the number of atomic energy levels about ten thousand.

Due to the large number of atomic energy levels required for a realistic simulation, JEKYLL is memory intensive, in particular if the Markov-chain solution to the MC state-machine (Sect.~\ref{s_markov_chains}) is used. While the MC packets as well as the processing of them are distributed over the available processes, only the processing of the grid cells is distributed. The reason for this is that during the radiative transfer, the MC packets potentially require access to all grid cells. This may be improved in future versions, but currently a typical model like the Type IIb model presented in Sect.~\ref{s_application} requires about 1 GB of memory per cell and node, which limits the total number of cells that can be used in a simulation.

JEKYLL uses several methods to reduce the computational effort. The most important are the Markov-Chain solution to the MC state-machine (Sect.~\ref{s_markov_chains}) and the use of a diffusion solver in the inner region (Sect.~\ref{s_diffusion_solver}). The speed-up achieved from the Markov-Chain solution to the MC state-machine is considerable and is discussed in Sect.~\ref{s_test_markov_chain}. However, as mentioned above, this method comes with the caveat of an increased use of memory resources. The speed-up achieved from the use of a diffusion solver in the inner region is essential for the code, and potentially huge at early times when the optical depths are large.

\section{Code comparisons}
\label{s_comp}

In this section we compare JEKYLL to ARTIS (\citetalias{Sim07} and \citetalias{Kro09}), SUMO (\citetalias{Jer11} and \citetalias{Jer12}), and CMFGEN \citepalias{Hil98}, three codes which have similar, but not identical capabilities. ARTIS provides a good test of the time-dependent MC radiative transfer, which is very similar, but only supports partial NLTE. SUMO on the other hand, provides a good test of the full NLTE problem, but requires steady-state, so no test of the time-dependent MC radiative transfer is possible. However, CMFGEN, which is similar to JEKYLL in physical assumptions but different in technique, does provide a test of the full time-dependent NLTE problem. In particular, as it solves the coupled radiation-matter problem and does not rely on $\Lambda$-iterations, it provides a way to show that the Lucy method actually converges to the correct solution. Here we present a comparison for a somewhat simplified test case, which still provides a good test of the full time-dependent NLTE problem. The comparisons to ARTIS, SUMO and CMFGEN are complementary, and taken together they provide a thorough test of the JEKYLL code.

\subsection{Comparison with ARTIS}
\label{s_comp_artis}

\begin{figure}[tbp!]
\includegraphics[width=0.5\textwidth,angle=0]{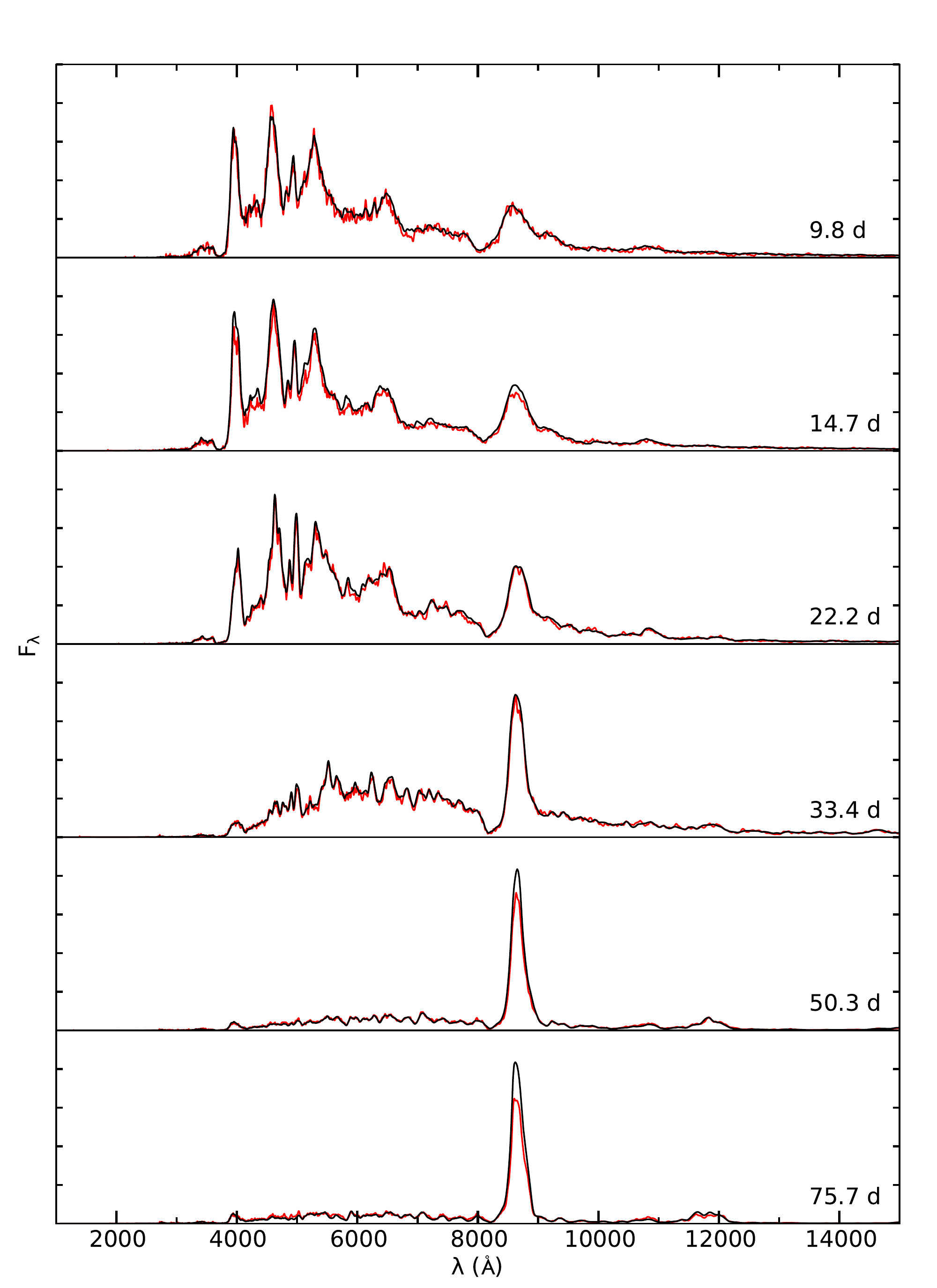}
\caption{Comparison of spectral evolution for model 12C as calculated with JEKYLL (black) and ARTIS (red). For this comparison both codes use LTE estimates for the population and the ionization state of the gas.}
\label{f_artis_comp_spec_evo}
\end{figure}

\begin{figure}[tbp!]
\includegraphics[width=0.5\textwidth,angle=0]{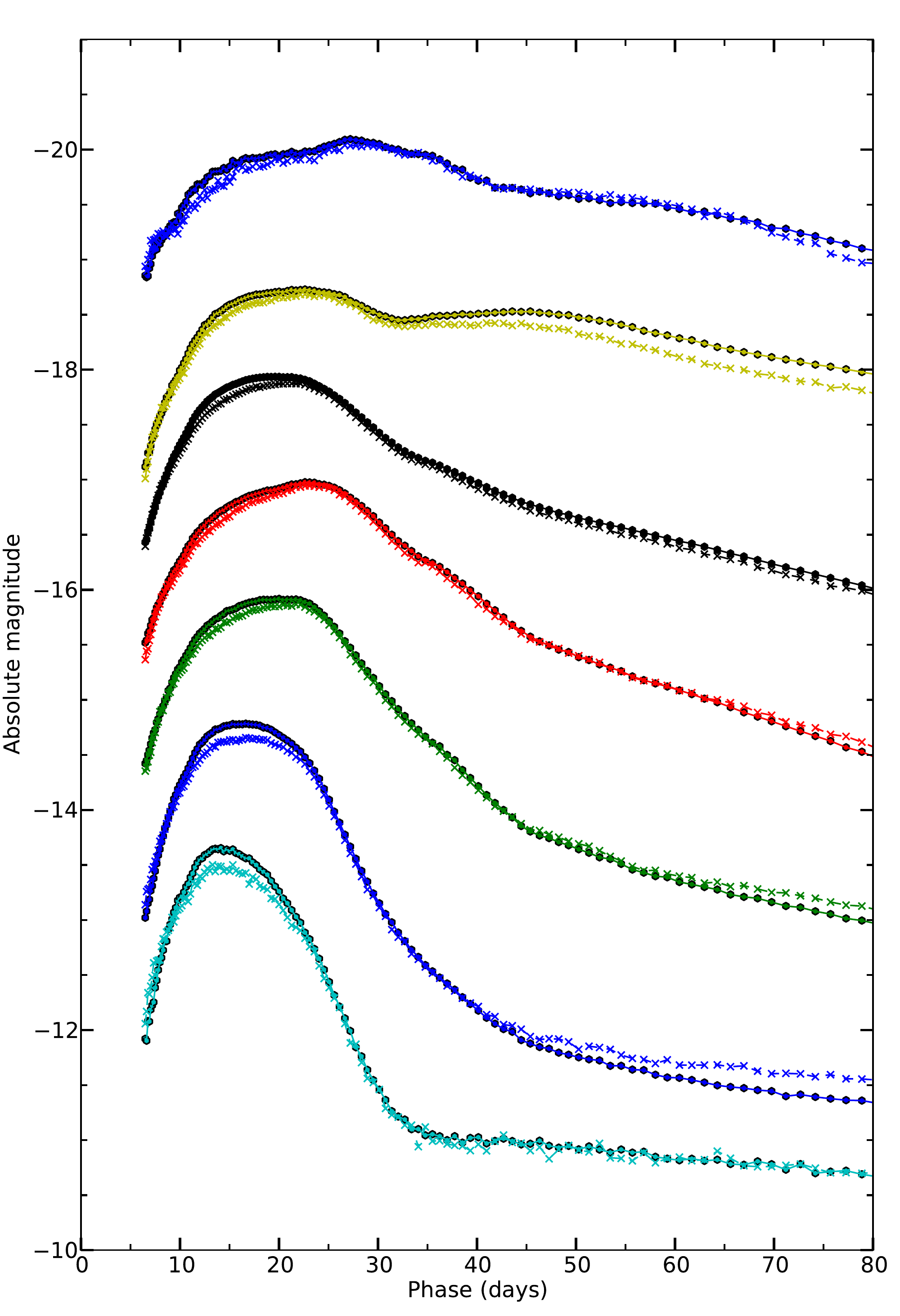}
\caption{Comparison of broad-band and bolometric lightcurves for model 12C as calculated with JEKYLL (solid lines and circles) and ARTIS (dashed lines and crosses). From bottom to top we show the U (cyan), B (blue), V (green), R (red), bolometric (black), I (yellow) and J (blue) lightcurves, which for clarity have been shifted with 2.0, 2.0, 1.5, 0.5, -1.0, -1.0 and -3.0 mags, respectively}
\label{f_artis_comp_lc_evo}
\end{figure}

\begin{figure}[tbp!]
\includegraphics[width=0.5\textwidth,angle=0]{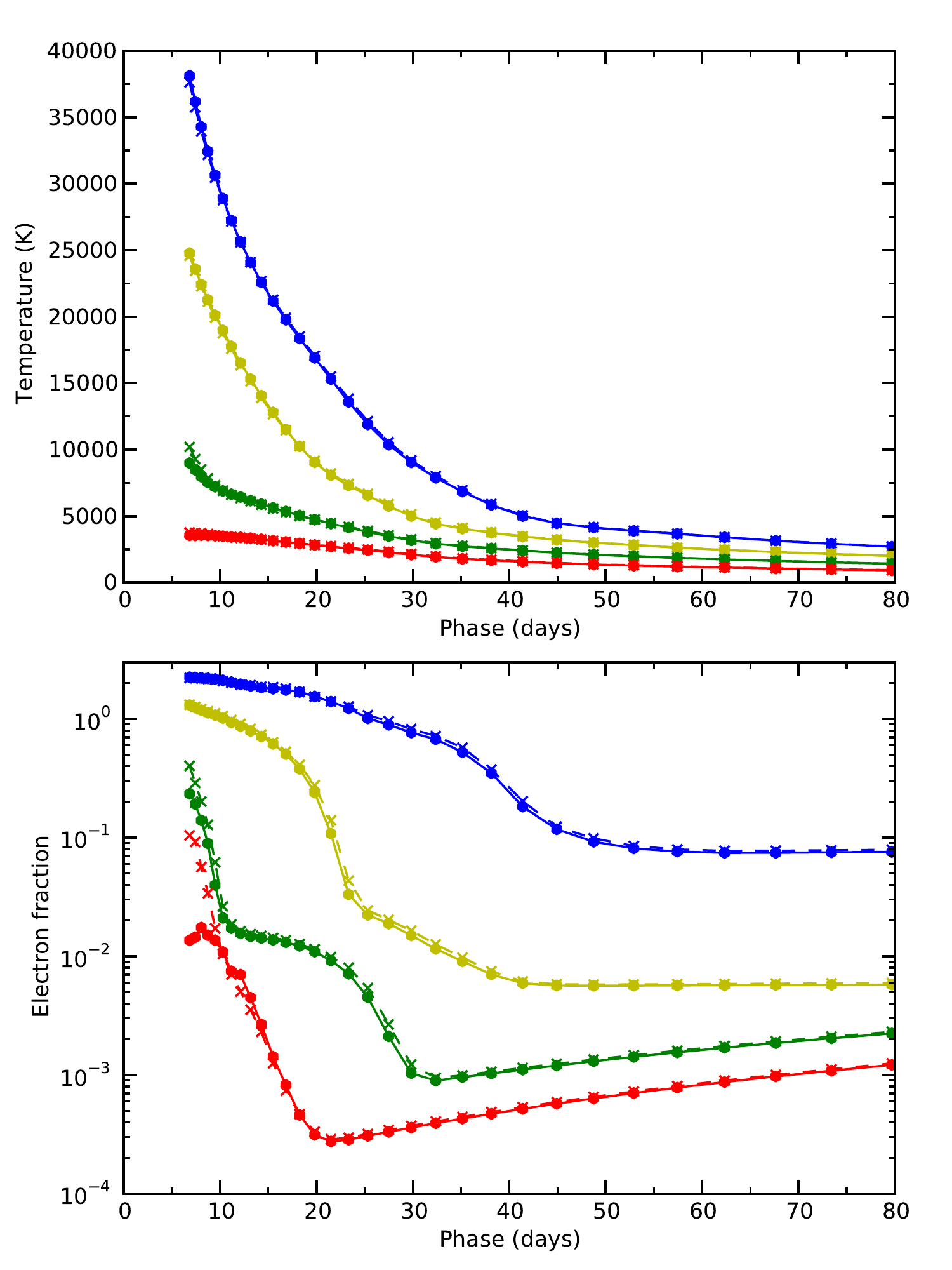}
\caption{Comparison of the evolution of the temperature (upper panel) and electron fraction (lower panel) in the oxygen core (blue), inner/outer (yellow/green) helium envelope and the hydrogen envelope (red) for model 12C as calculated with JEKYLL (circles and solid lines) and ARTIS (crosses and dashed lines).}
\label{f_artis_comp_matter_evo}
\end{figure}

ARTIS is a spectral synthesis code aimed for the photospheric phase presented in \citetalias{Sim07} and \citetalias{Kro09}. Both ARTIS and JEKYLL are based on the Lucy method, but ARTIS only supports a simplified NLTE treatment\footnote{A more general NLTE treatment and the inclusion of non-thermal processes in ARTIS are currently under development.}, where the excited states are populated according to LTE and the energy deposited by the radioactive decays goes solely into heating. On the other hand, the current version of JEKYLL assumes a spherical symmetric geometry, which is not a limitation in ARTIS. In addition, ARTIS calculates the deposition of the radioactive decay energy by Compton scattering, photo-electric absorption and pair production, whereas JEKYLL uses effective grey opacities (based on such calculations). There are also differences in the NLTE ionization treatment, in particular with respect to the calculation of photo-ionization rates, and due to this we decided to run ARTIS in its LTE mode. This still allows for a complete test of the time-dependent MC radiative transfer, which is the main purpose of the ARTIS comparison.

For the comparison we have used the Type IIb model 12C from \citetalias{Jer15}, which we also have used for the application to Type IIb SNe in Sect.~\ref{s_application}. The original model was converted to microscopically mixed form and re-sampled to a finer spatial grid as described in Sect.~\ref{s_model_description}. To synchronize JEKYLL with ARTIS, it was configured to run in time-dependent (radiative transfer) mode using the LTE solver, and the ARTIS atomic data was automatically converted to the JEKYLL format. The details of the code configurations and the atomic data used are given in Appendix~\ref{a_conf_data_comp_artis}, and we find the synchronization good enough for a meaningful comparison. We note, that as non-thermal processes are crucial for the population of the excited \ion{He}{i} states, the characteristic \ion{He}{i} signature of Type IIb SNe is not reproduced. 

In Figs.~\ref{f_artis_comp_spec_evo} and \ref{f_artis_comp_lc_evo} we compare the spectral evolution and the lightcurves, respectively, whereas in Fig.~\ref{f_artis_comp_matter_evo} we compare the evolution of the temperature and the electron fraction. As the grey approximation used in ARTIS (see Appendix~\ref{a_conf_data_comp_artis}) affects the early evolution, we compare the models after 6 days, although the effect seems to last for a few more days in some quantities (e.g.~the electron fraction). As can be seen, the general agreement is good in both the observed and the state quantities. The most conspicuous discrepancy appears in the~\ion{Ca}{ii} 8498,8542,8662 \AA~line after $\sim$40 days, and gives rise to a $\sim$15 percent discrepancy in the $I$-band lightcurve. Another discrepancy appears after $\sim$50 days in the $B$-band, growing towards $\sim$15 percent at 80 days. There is also a small (<5 percent) but clear difference in the bolometric tail luminosity, reflecting a similar difference in the radioactive energy deposition. This is due to the the more approximate method for this used by JEKYLL, which may also explain the differences in the tail broad-band lightcurves. There are also minor differences in the diffusion peak lightcurves, most pronounced in the $U$- and $B$-bands, which could be related to the simplified treatment at high optical depths in ARTIS and JEKYLL (grey approximation and diffusion solver, respectively; see Appendix~\ref{a_conf_data_comp_artis}). Summarizing, although there are some minor differences in the spectra and the lightcurves, we find the overall agreement to be good.

\subsection{Comparison with SUMO}
\label{s_comp_sumo}

\begin{figure}[tbp!]
\includegraphics[width=0.5\textwidth,angle=0]{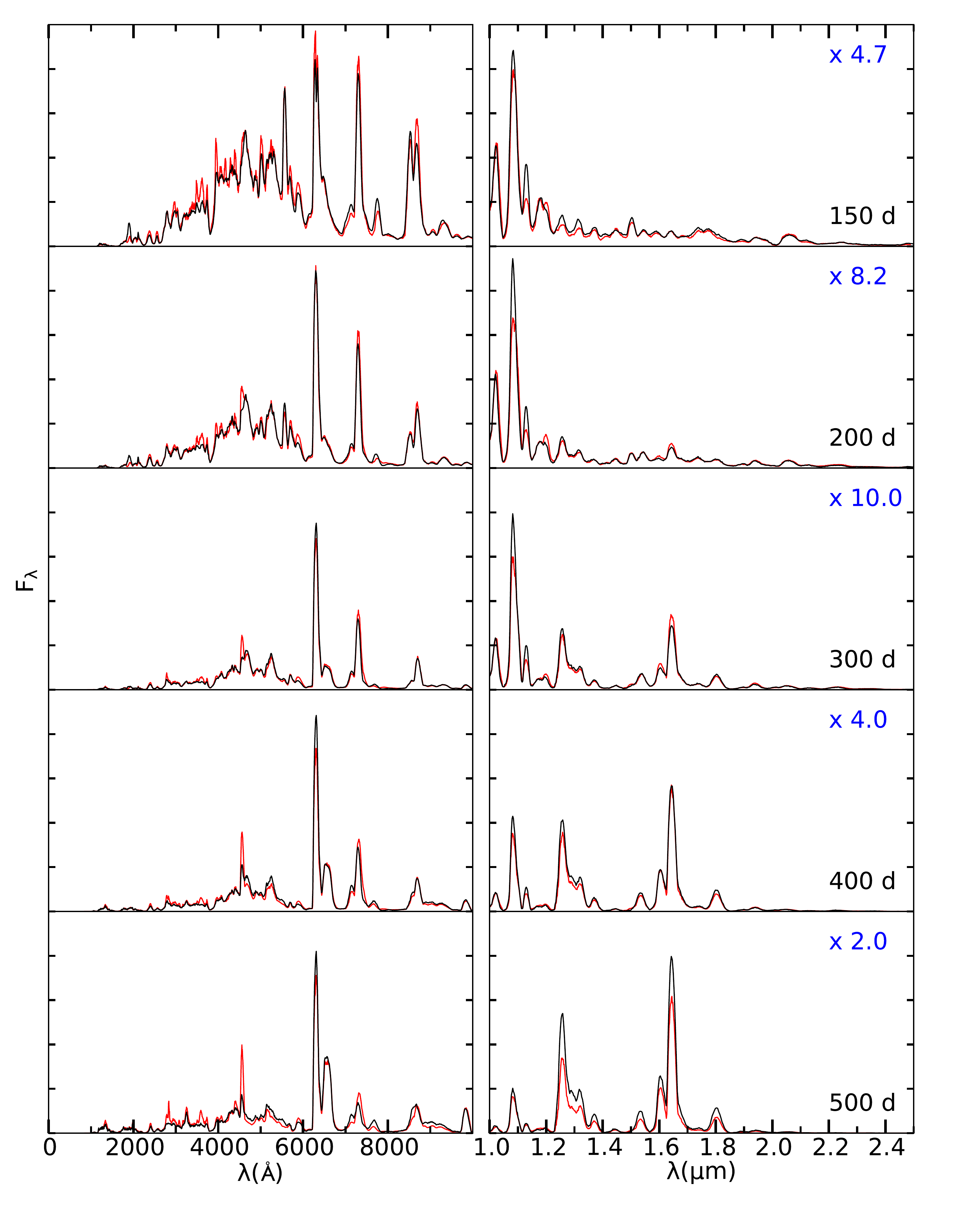}
\caption{Comparison of optical (left panel) and NIR (right panel) spectra for model 13G at 150, 200, 300, 400 and 500 days as calculated with JEKYLL (black) and SUMO (red). For clarity the NIR spectra have been scaled with respect to the optical spectra with the factor given in the upper right corner.}
\label{f_j14_comp_spec_evo}
\end{figure}

\begin{figure}[tbp!]
\includegraphics[width=0.5\textwidth,angle=0]{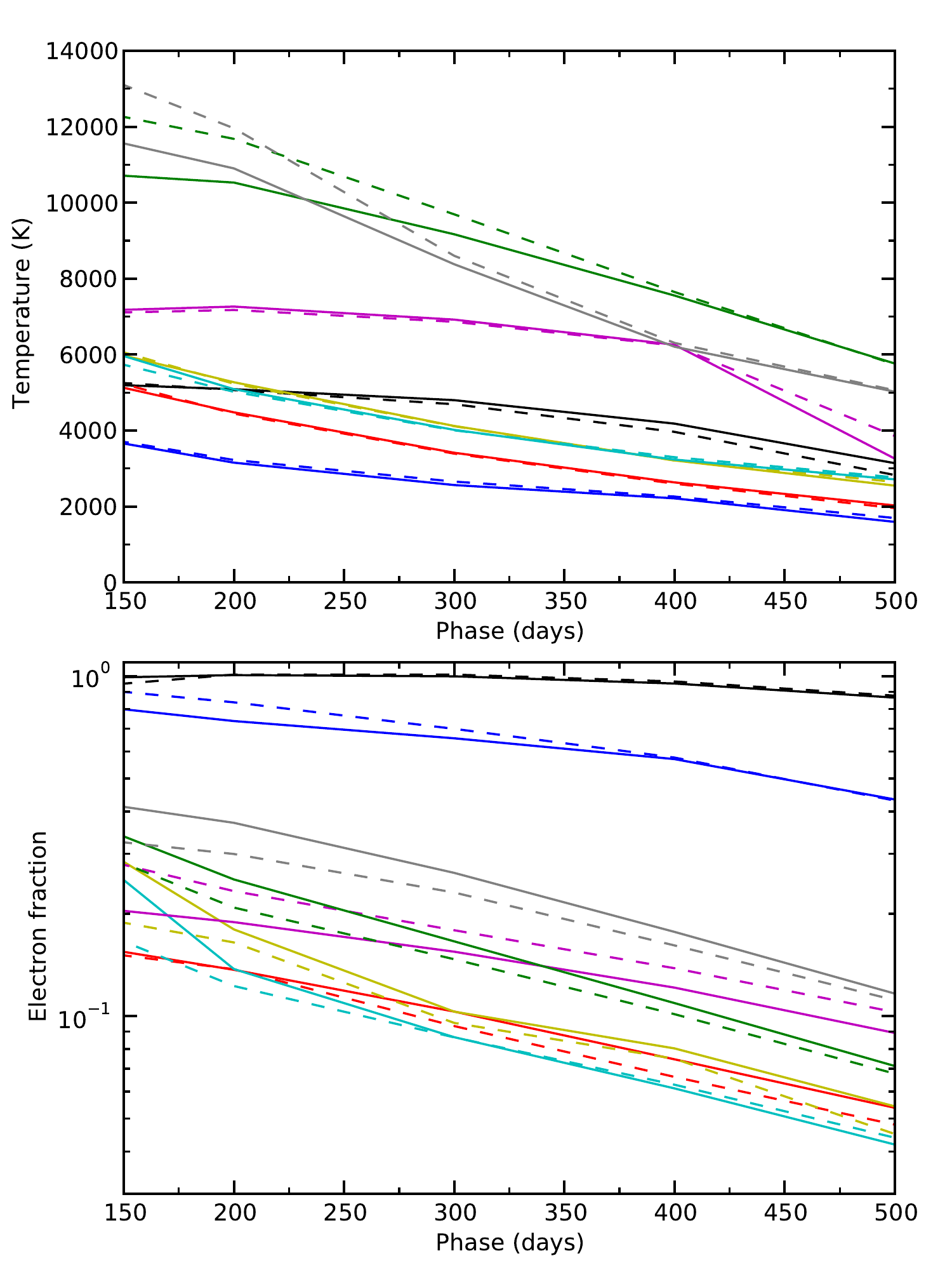}
\caption{Comparison of the evolution of the temperature (upper panel) and the electron fraction (lower panel) for model 13G in the Fe/Co/He (black), Si/S (blue), O/Si/S (red), O/Ne/Mg (yellow), O/C (cyan), He/C (magenta), He/N (green) and H (grey) zones as calculated with JEKYLL (solid lines) and SUMO (dashed lines).}
\label{f_j14_comp_matter_evo}
\end{figure}

\begin{figure}[tbp!]
\includegraphics[width=0.5\textwidth,angle=0]{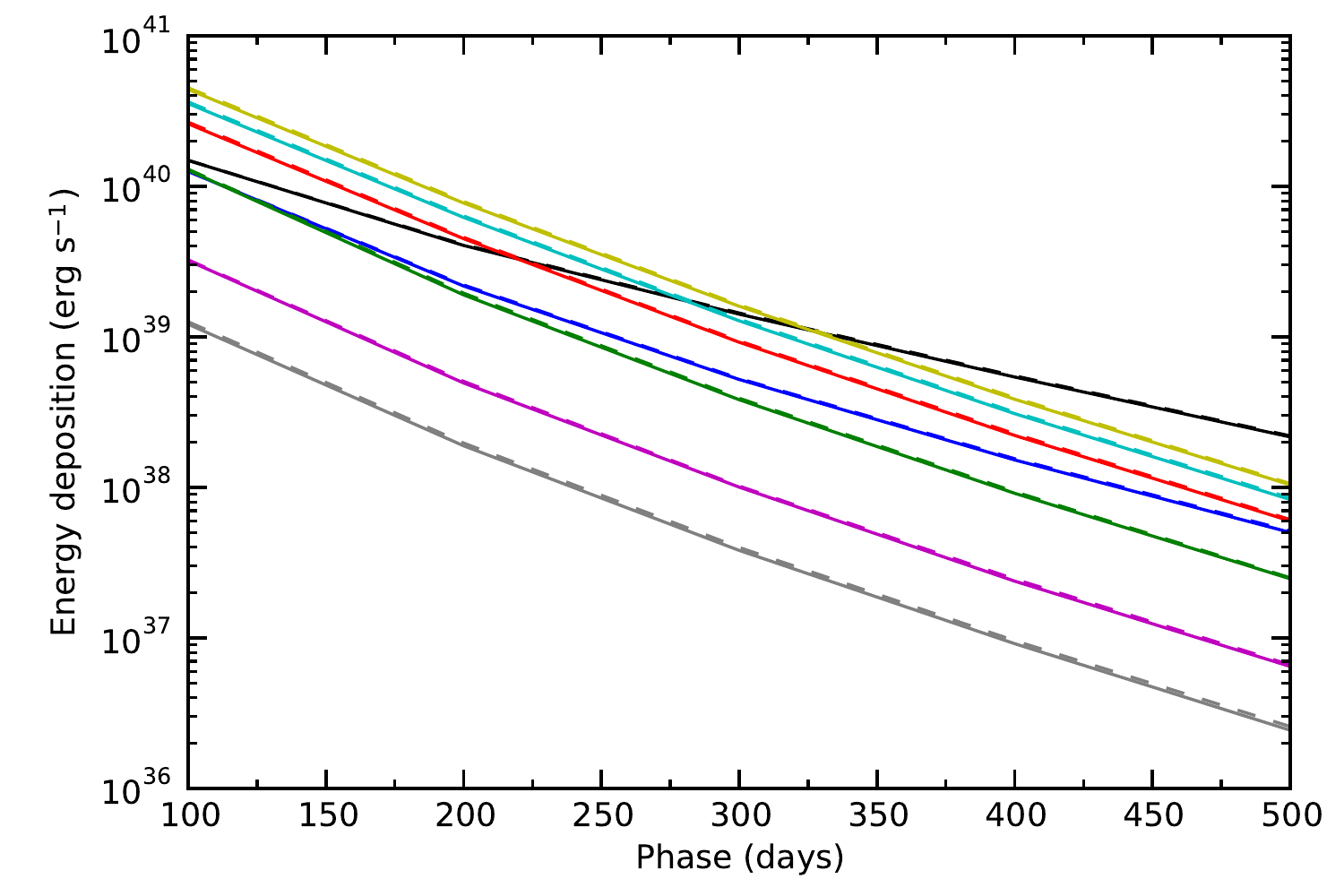}
\caption{Comparison of the evolution of the radioactive energy deposition for model 13G in the Fe/Co/He (black), Si/S (blue), O/Si/S (red), O/Ne/Mg (yellow), O/C (cyan), He/C (magenta), He/N (green) and H (grey) zones as calculated with JEKYLL (solid lines) and SUMO (dashed lines).}
\label{f_j14_comp_lradio_evo}
\end{figure}

SUMO is a spectral synthesis code aimed for the nebular phase presented in \citetalias{Jer11} and \citetalias{Jer12}. Similar to JEKYLL, it uses a $\Lambda$-iteration scheme, where the radiative transfer is solved with a MC method and the state of the matter determined from statistical and thermal equilibrium. Except for the steady-state assumption (for the radiative transfer), which is required by SUMO and an option in JEKYLL, the main difference between SUMO and JEKYLL is the MC technique used. Whereas JEKYLL is based on the Lucy method, where conservation of packet energy is enforced, SUMO uses another approach. Except for electron scattering and excitations to high lying states, the packet energy absorbed in free-free, bound-free and bound-bound processes is not re-emitted. As long as these processes are included in the emissivity from which the packets are sampled, this gives the correct solution in the limit of convergence. However, it could be an issue for the rate of convergence,  in particular at high absorption depths, and the method is probably not suited for the photospheric phase. There are also a few differences in the physical assumptions. Whereas JEKYLL correctly samples the frequency dependence of the bound-free emissivity, this is done in a simplified manner for all species but hydrogen by SUMO. On the other hand, JEKYLL does not take the escape probability from continua and other lines in the Sobolev resonance region into account. Yet another difference is that SUMO does not include the highest lying states in the NLTE solution \citep[see][]{Jer12}. However, in general the physical assumptions are similar.

For the comparison we have used the Type IIb model 13G from \citetalias{Jer15}, and have run models with JEKYLL at 150, 200, 300, 400 and 500 days. To synchronize JEKYLL with SUMO, it was configured to run in steady-state mode using the NLTE solver, and we have tried to synchronize the atomic data as much as possible. The details of the code configurations and the atomic data used are given in Appendix~\ref{a_conf_data_comp_sumo}, and although not complete, we find the synchronization good enough for a meaningful comparison.

A comparison of the spectral evolution is shown in Fig.~\ref{f_j14_comp_spec_evo}, and in Figs.~\ref{f_j14_comp_matter_evo} and \ref{f_j14_comp_lradio_evo} we compare the evolution of the temperature, the electron fraction and the radioactive energy deposition in each of the different nuclear burning zones (see \citetalias{Jer15}). As can be seen, the general agreement of the spectra is quite good, although the match is slightly worse at 500 days. The largest discrepancies are seen in the \ion{Mg}{i}] 4571 \AA~line, the \ion{O}{i} 11290,11300 \AA~line before 300 days, the \ion{He}{i} 10830 \AA~line at 200-300 days, and a number of features originating from the Fe/Co/He zone at 500 days. That one of the largest discrepancies is seen in the \ion{Mg}{i}] 4571 \AA~line is not surprising as magnesium is mainly ionized and the \ion{Mg}{i} fraction is small (see \citetalias{Jer15}). This makes the strength of the \ion{Mg}{i}] 4571 \AA~line sensitive to this fraction, in turn sensitive to the network of charge transfer reactions. 

The evolution of the temperature shows a good agreement and the differences are mainly below $\sim$5 percent. An exception is the He/N and H zones at early times, and in particular at 150 days where the difference is $\sim$15 percent. The evolution of the electron fraction shows a worse agreement, but the differences are mainly below $\sim$10 percent. Again, the agreement is worst at early times, and in particular at 150 days when the electron fractions in the O/Ne/Mg and O/C zones differ by $\sim$30 percent. This discrepancy is reflected in for example the \ion{O}{i} 11290,11300 \AA~line discussed above, but in general the spectral agreement at 150 days is quite good. 

The evolution of the radioactive energy deposition shows an excellent agreement. This shows that the radiative transfer of the $\gamma$-packets (Sect.~\ref{s_packets} and \ref{s_packet_propagation}), representing the $\gamma$-rays (and leptons) emitted in the radioactive decays, as well as the method for macroscopic mixing of the material (Sect.~\ref{s_virtual_cells} and \ref{s_packet_propagation}), works as intended. Summarizing, although there are some notable differences both in the spectra and the state variables, we find the overall agreement to be good, in particular as the data and the methods are not entirely synchronized.

\subsection{Comparison with CMFGEN}
\label{s_comp_cmfgen}

CMFGEN is a general purpose spectral synthesis code presented in its steady-state version in \citetalias{Hil98}, and extended with time-dependence in \citet{Des08a,Des10} and \citet{Hil12} and non-thermal processes in \citet{Des12}. It is similar to JEKYLL in the physical assumptions, but uses a different method to solve the NLTE problem, where the coupled system of differential equations for the matter and the radiation field is solved by a linearization technique. The potential difficulties with convergence in $\Lambda$-iteration based methods (Sect.~\ref{s_lambda_iter}) are therefore avoided, and the comparison provides a good test of the convergence properties of the $\Lambda$-iteration and MC based method used in JEKYLL (i.e.~the one by Lucy). The main difference in the physical assumptions is that JEKYLL assumes steady state for the matter, whereas CMFGEN does not. In addition, CMFGEN does not rely on the Sobolev approximation, but this is likely of less importance at the high velocity gradients present in SN ejecta. We note, however, that in the Sobolev approximation, absorption in continua and other lines within the resonance region is ignored, which is a potential problem. On the other hand, the choice of micro-turbulent velocity in CMFGEN might lead to an overestimate of the overlap between lines.

To synchronize with JEKYLL, time-dependence for the matter needs to be switched off in CMFGEN. However, as this turned out to be difficult, we have instead added support for limited time-dependence in JEKYLL. This was done by adding an option to use the more general NLTE rate-equations, which was achieved by adding the time-derivative 
\begin{equation}
\rho {{D(n_{I,j}/\rho)} \over {Dt}}
\end{equation}
to the right-hand side of Eq.~\ref{eq_stat_equi} (see \citealt{Des08a}). This accounts for the effect of time-dependence on the degree of ionization, which is the most important one, at least in the test model. We note, that time-dependence is only added in this limited form to facilitate the comparison with CMFGEN, and is not explored further in the paper. A general upgrade of JEKYLL to full time-dependence will be presented in a forth-coming paper.

For the comparison we have used a model of a red supergiant of 15 M$_\odot$ initial mass, evolved with MESA \citep{Pax11,Pax13} and exploded with an energy of 1 Bethe with the hydrodynamical code HYDE \citepalias{Erg14}. The JEKYLL and CMFGEN simulations begin at 25 days, and for the test we use a simplified composition consisting of hydrogen, helium, oxygen and calcium. 
To synchronize, the CMFGEN atomic data was automatically converted to the JEKYLL format, and the details of the code configurations and the atomic data used are given in Appendix~\ref{a_conf_data_comp_cmfgen}, 

\begin{figure}[tbp!]
\includegraphics[width=0.5\textwidth,angle=0]{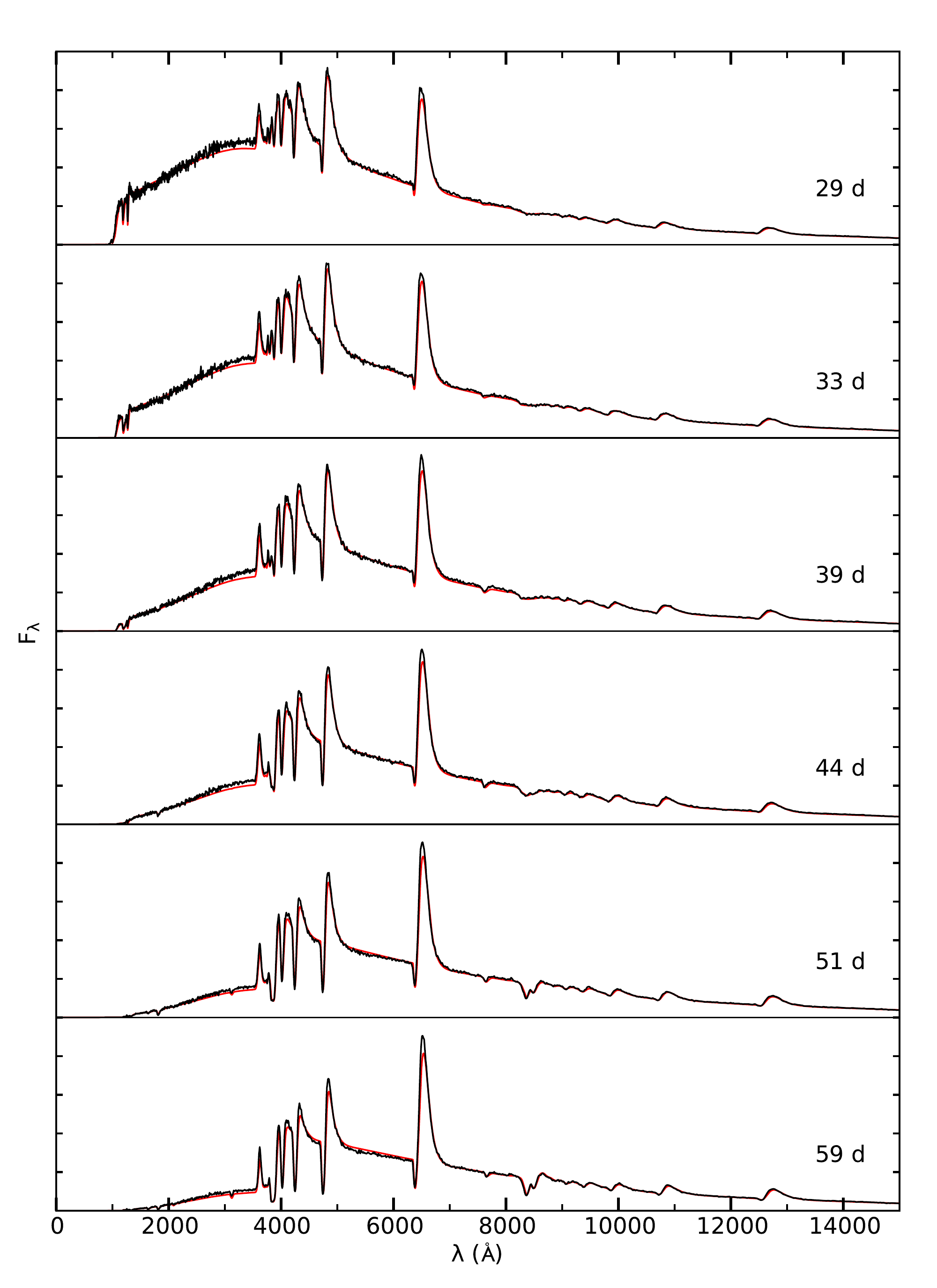}
\caption{Comparison of spectral evolution for the test model as calculated with JEKYLL (black) and CMFGEN (red).}
\label{f_cmfgen_comp_spec_evo}
\end{figure}

\begin{figure}[tbp!]
\includegraphics[width=0.5\textwidth,angle=0]{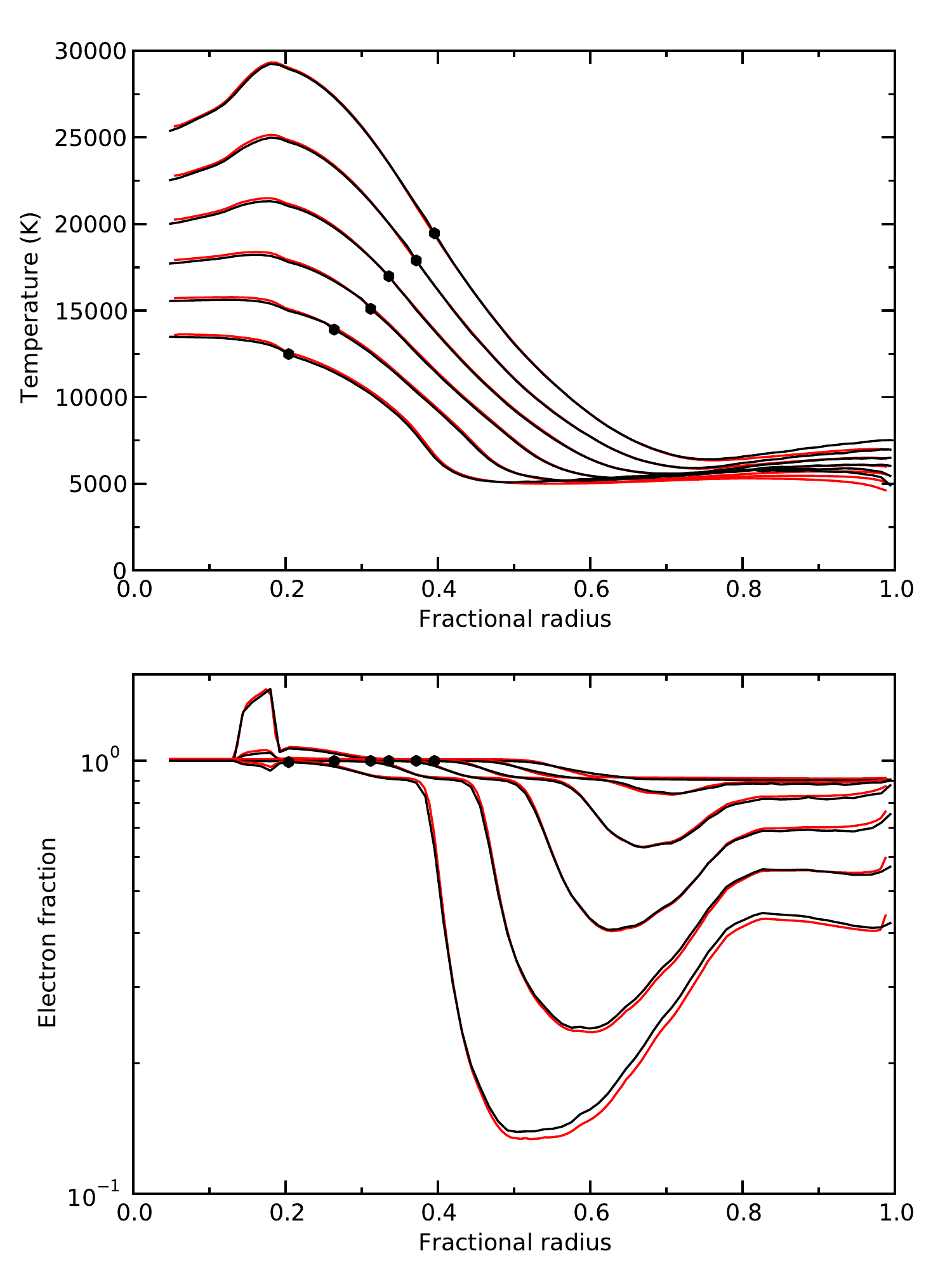}
\caption{Comparison of the evolution of the temperature (upper panel) and electron fraction (lower panel) at the same epochs as in Fig.~\ref{f_cmfgen_comp_spec_evo} for the test model as calculated with JEKYLL (black) and CMFGEN (red). The border between the regions handled by the diffusion solver and the MC radiative transfer solver have been marked with black circles.}
\label{f_cmfgen_comp_matter_evo}
\end{figure}

A comparison of the spectral evolution is shown in Fig.~\ref{f_cmfgen_comp_spec_evo}, and in Fig.~\ref{f_cmfgen_comp_matter_evo} we compare the evolution of the temperature and the electron fraction. As can be seen, the overall agreement is good in both the spectra and the matter quantities. The largest differences in the spectra are a somewhat higher flux in the Balmer continuum and a bit stronger emission in the Balmer lines in the JEKYLL model. The electron fraction is in good agreement, but the temperature is slightly higher in the outer region in the JEKYLL model. Given that time-dependence is only partly implemented in JEKYLL, and is missing in the thermal energy equation, differences at this level are not surprising. 

The good overall agreement found in both the spectra and the matter quantities shows that the $\Lambda$-iteration and MC based method used in JEKYLL (i.e.~the one by Lucy) does indeed converge to a solution close to the correct one\footnote{CMFGEN may not be free of bugs, and may also have other short-comings, so this has to be taken with a grain of salt. In addition, physics that might be required to correctly model SN ejecta (as e.g.~3-D) are still missing in both CMFGEN and JEKYLL.}, at least in the specific case tested here. Departures from LTE are large (typically a factor of ten or larger) in the optically thin region, so although based on a model with simplified composition, the comparison provides a good test of the time-dependent NLTE capabilities.

\section{Application and tests}
\label{s_application}

In this section we provide an example of a time-dependent NLTE model based on a fully realistic ejecta model. The ejecta model (12C) has been taken from the set of Type IIb models constructed by \citetalias{Jer15}, and was found to give the best match to the observed nebular spectra \citepalias{Jer15} and lightcurves \citepalias{Erg15} of SN 2011dh. For a more detailed discussion of the model and a comparison to SN 2011dh we refer to Paper 2, where we also explore other models.

In addition to a brief discussion of the model and its evolution, we investigate the effect of NLTE on the spectra and the lightcurves, in particular with respect to non-thermal ionization and excitation. Based on the model, we also investigate the convergence of the $\Lambda$-iterations, and provide tests of our most important (technical) extensions to the original Lucy method, that is the use of a diffusion solver in the inner region, the packet sampling control and the Markov-chain solution to the MC state-machine.

\subsection{Ejecta model}
\label{s_model_description}

A full description of the Type IIb model 12C is given in \citetalias{Jer15}, but we summarize the basic properties here. It is based on a model by \citet{Woo07} with an initial mass of 12 M$_\odot$, from which we have taken the masses and abundances for the carbon-oxygen core and the helium envelope. We have assumed the carbon-oxygen core to be fully mixed and to have a constant (average) density, and the helium envelope to have the same density profile as the best-fit model for SN 2011dh by \citet{Ber12}. In addition, a 0.1 M$_\odot$ hydrogen envelope based on models by \citet{Woo94} has been attached. We note, that the ejecta model explored here is a microscopically mixed version, in which the abundances in the nuclear burning zones (see \citetalias{Jer15}) have been averaged. For the macroscopically mixed version we refer to Paper 2, where we also discuss the effect of this difference on the model evolution. To be suitable for modelling in the photospheric phase, the original ejecta model has also been re-sampled to a finer spatial grid.

\subsection{Model evolution}
\label{s_model_evolution}

\begin{figure}[tbp!]
\includegraphics[width=0.5\textwidth,angle=0]{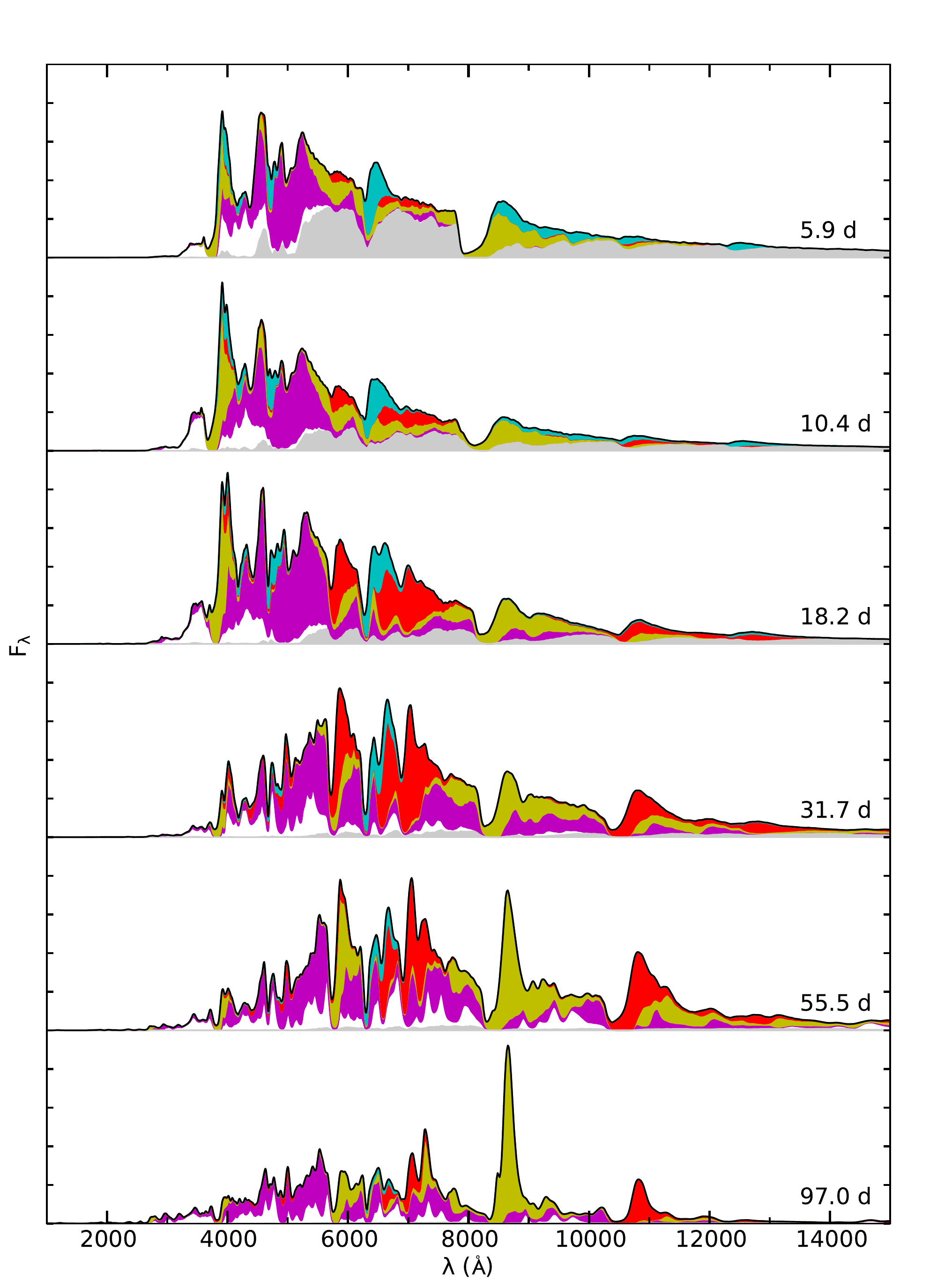}
\caption{Spectral evolution for model 12C as calculated with JEKYLL. In the spectra we show the contributions to the emission from bound-bound transitions of hydrogen (cyan), helium (red), carbon-calcium (yellow), scandium-manganese (white) and iron-nickel (magenta) as well as continuum processes (grey).}
\label{f_12C_spec_trans_evo}
\end{figure}

\begin{figure}[tbp!]
\includegraphics[width=0.5\textwidth,angle=0]{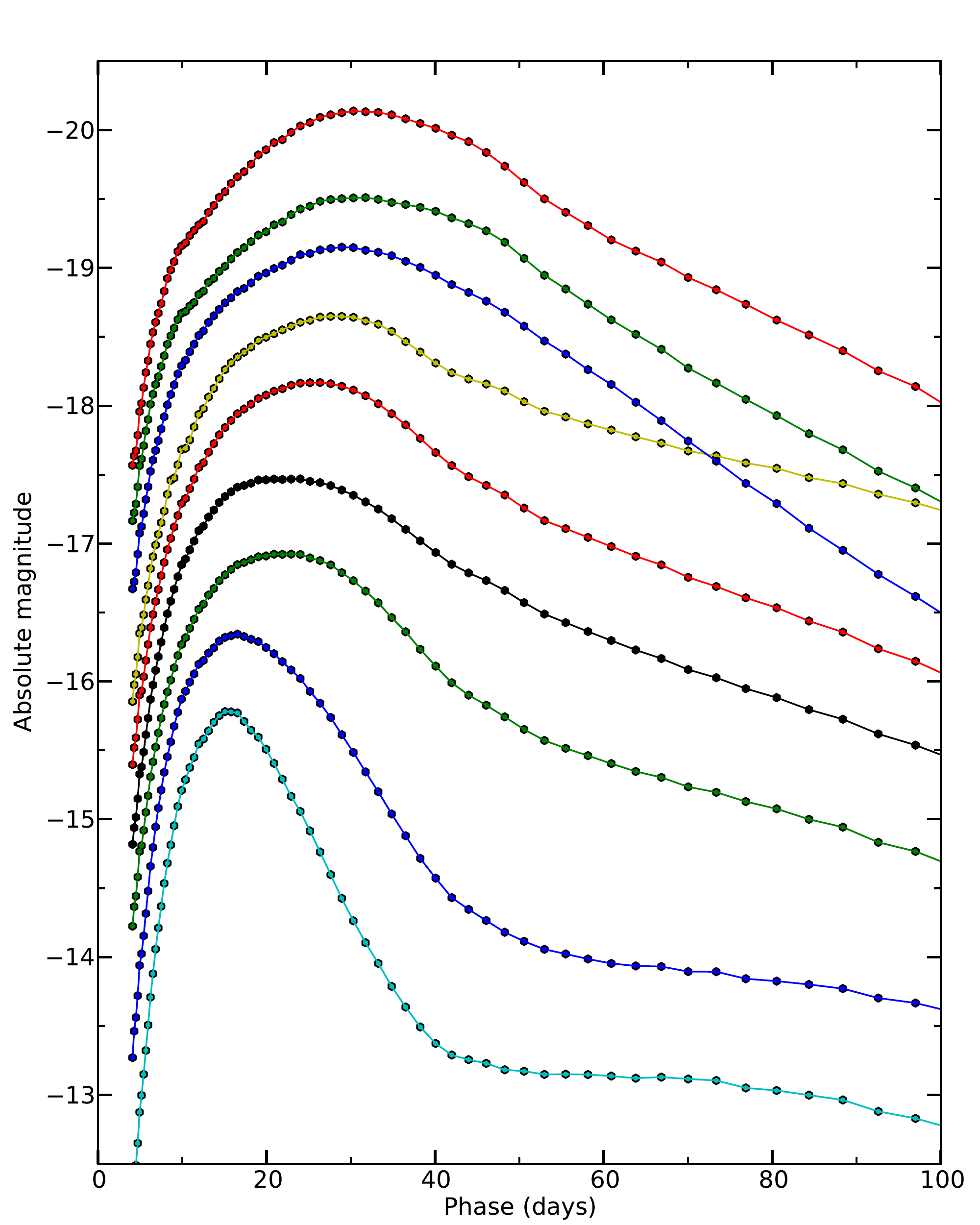}
\caption{Broad-band and bolometric lightcurves for model 12C as calculated with JEKYLL . From bottom to top we show the U (cyan), B (blue), V (green), R (red), bolometric (black), I (yellow), J (blue), H (green) and K (red) lightcurves, which for clarity have been shifted with 0.4, 0.0, -0.9, 0.0, -0.9, -1.4, -1.7, -2.1 and -2.4 mags, respectively.}
\label{f_12C_lightcurve}
\end{figure}

\begin{figure}[tbp!]
\includegraphics[width=0.5\textwidth,angle=0]{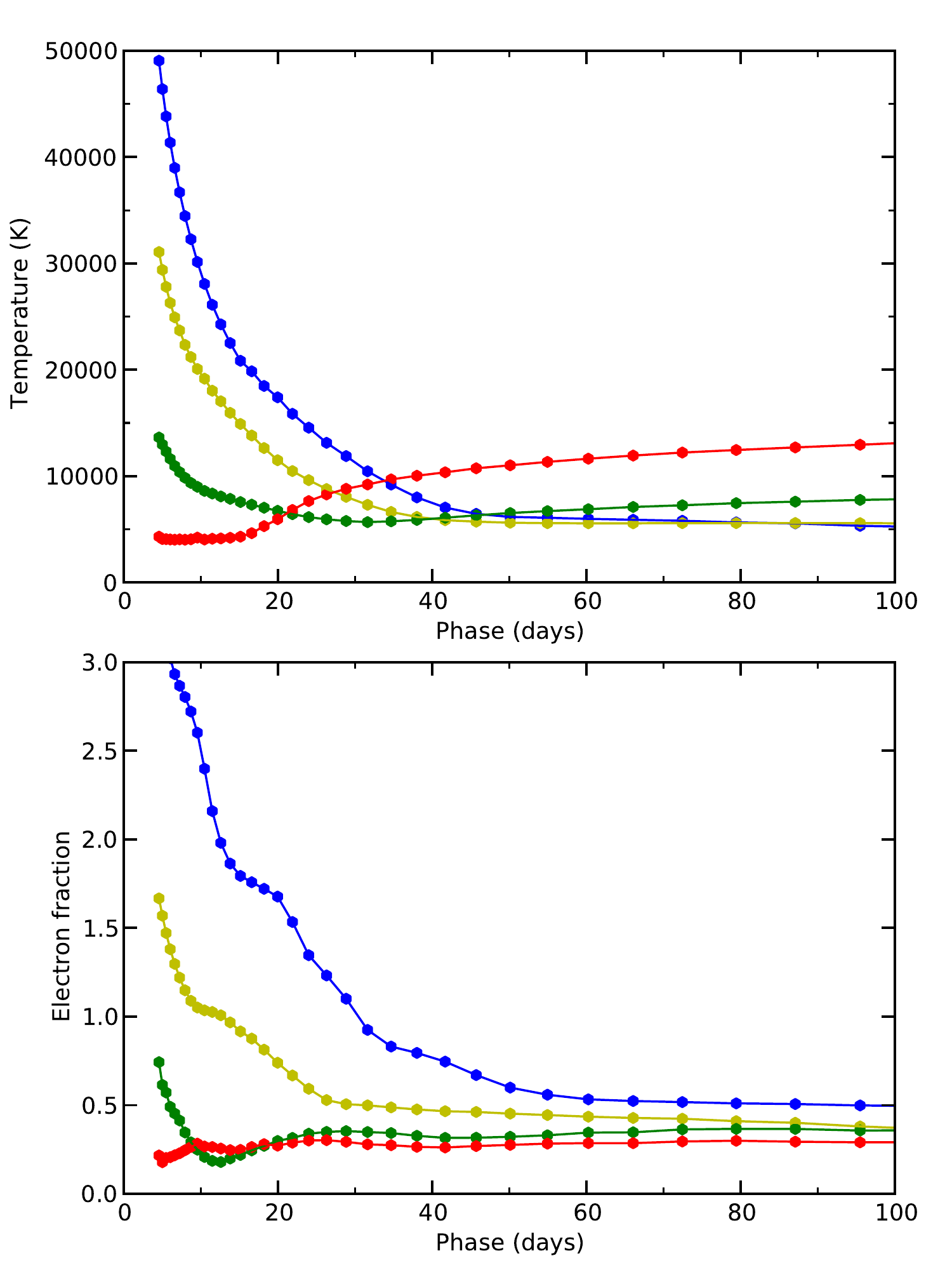}
\caption{Evolution of the temperature (upper panel) and electron fraction (lower panel) in the oxygen core (blue), inner/outer (yellow/green) helium envelope and the hydrogen envelope (red) for model 12C.}
\label{f_12C_matter_evo}
\end{figure}

JEKYLL was configured to run in time-dependent (radiative transfer) mode, using the NLTE solver based on an updated version of the \citetalias{Jer15} atomic data, and we give the details of the configuration and the atomic data in Appendix \ref{a_conf_data_app}. The model was evolved from 1 to 100 days, and the initial temperature profile was taken from the best-fit model for SN 2011dh from \citetalias{Erg15}. Figure~\ref{f_12C_spec_trans_evo} shows the spectral evolution, whereas Fig.~\ref{f_12C_lightcurve} shows the lightcurves and Fig.~\ref{f_12C_matter_evo} the evolution of the temperature and the electron fraction. In Fig.~\ref{f_12C_spec_trans_evo} we also display the process giving rise to the emission, based on the last emission events for the MC packets (excluding electron scattering).

The main signature of a Type IIb SN is the transition from a hydrogen to a helium dominated spectrum, and this is well reproduced by the model. Initially, the hydrogen lines are strong and emission from the hydrogen envelope is dominating. Between 10 and 15 days the helium lines appear, grow stronger, and eventually dominate the spectrum at $\sim$40 days. Hydrogen line emission disappears on a similar time-scale, completing the transition, although the Balmer lines remain considerably longer in absorption. After $\sim$40 days the carbon-oxygen core gets increasingly transparent and the amount of realism in the microscopically mixed model starts to degrade, exemplified by the strong calcium NIR triplet at 100 days. As is demonstrated in Paper 2, the macroscopically mixed version of the model does a considerable better job in reproducing observations in the nebular phase.

The lightcurves show the characteristic bell shape of stripped-envelope (SE; Type IIb, Ib and Ic) SNe, and as discussed in Paper 2, their change in shape with effective wavelength (as e.g.~a broader peak for redder bands) is in good agreement with observational studies. It is worth noting that the behaviour of model 12C is similar to that of the NLTE models of SE SNe presented by \citet{Des15,Des16}. Those models were evolved with CMFGEN, and in particular the Type IIb model 3p65Ax1 shares many properties with model 12C.

\subsection{The effect of NLTE}
\label{s_effect_nlte}

\begin{figure}[tbp!]
\includegraphics[width=0.5\textwidth,angle=0]{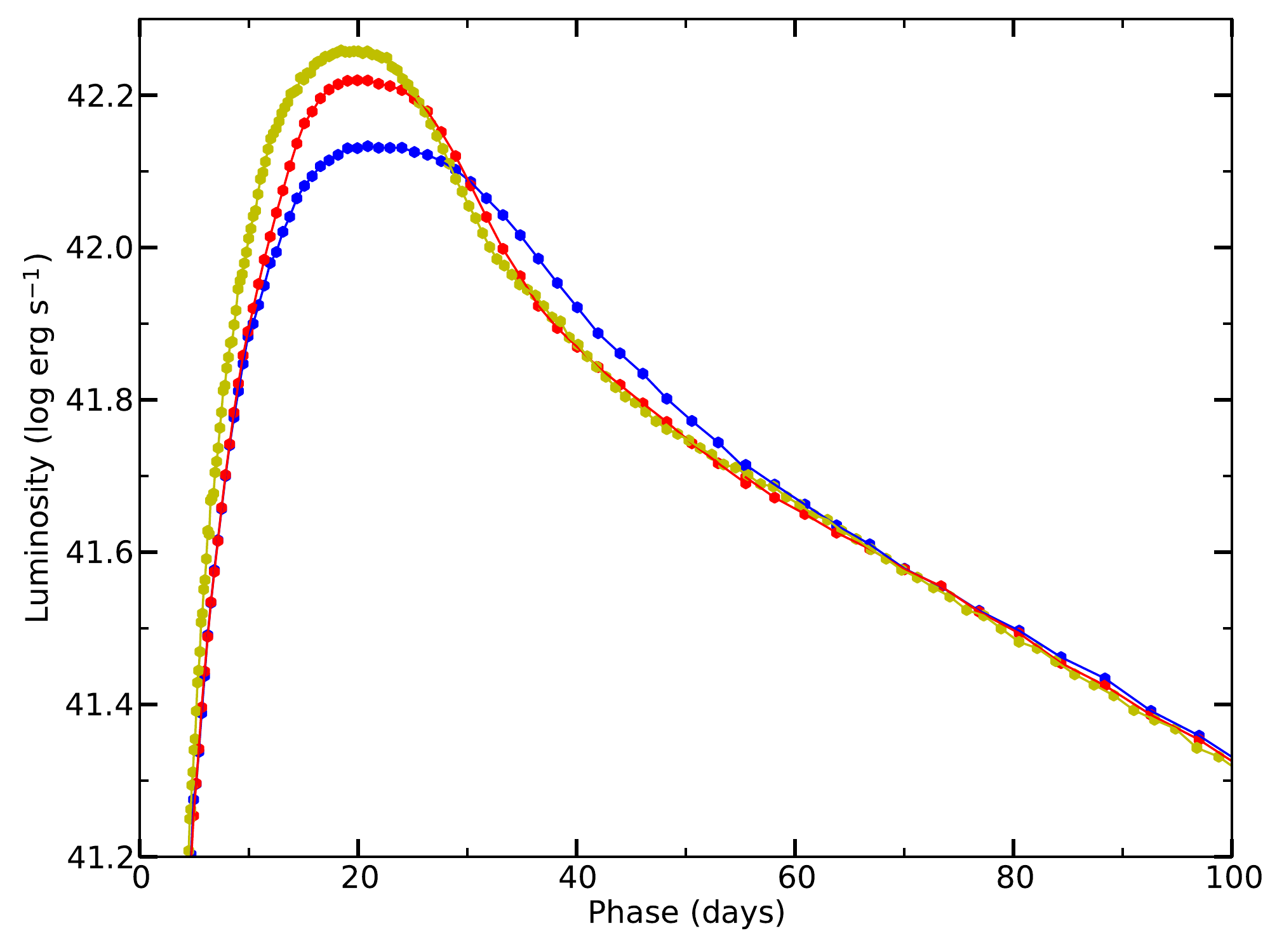}
\caption{Bolometric lightcurve for model 12C calculated with (blue) and without (red) non-thermal ionization and excitation. We also show the bolometric lightcurve for the LTE version of model 12C (yellow) presented in Sect.~\ref{s_comp_artis}.}
\label{f_12C_nlte_comp_lc_evo}
\end{figure}

\begin{figure}[tbp!]
\includegraphics[width=0.5\textwidth,angle=0]{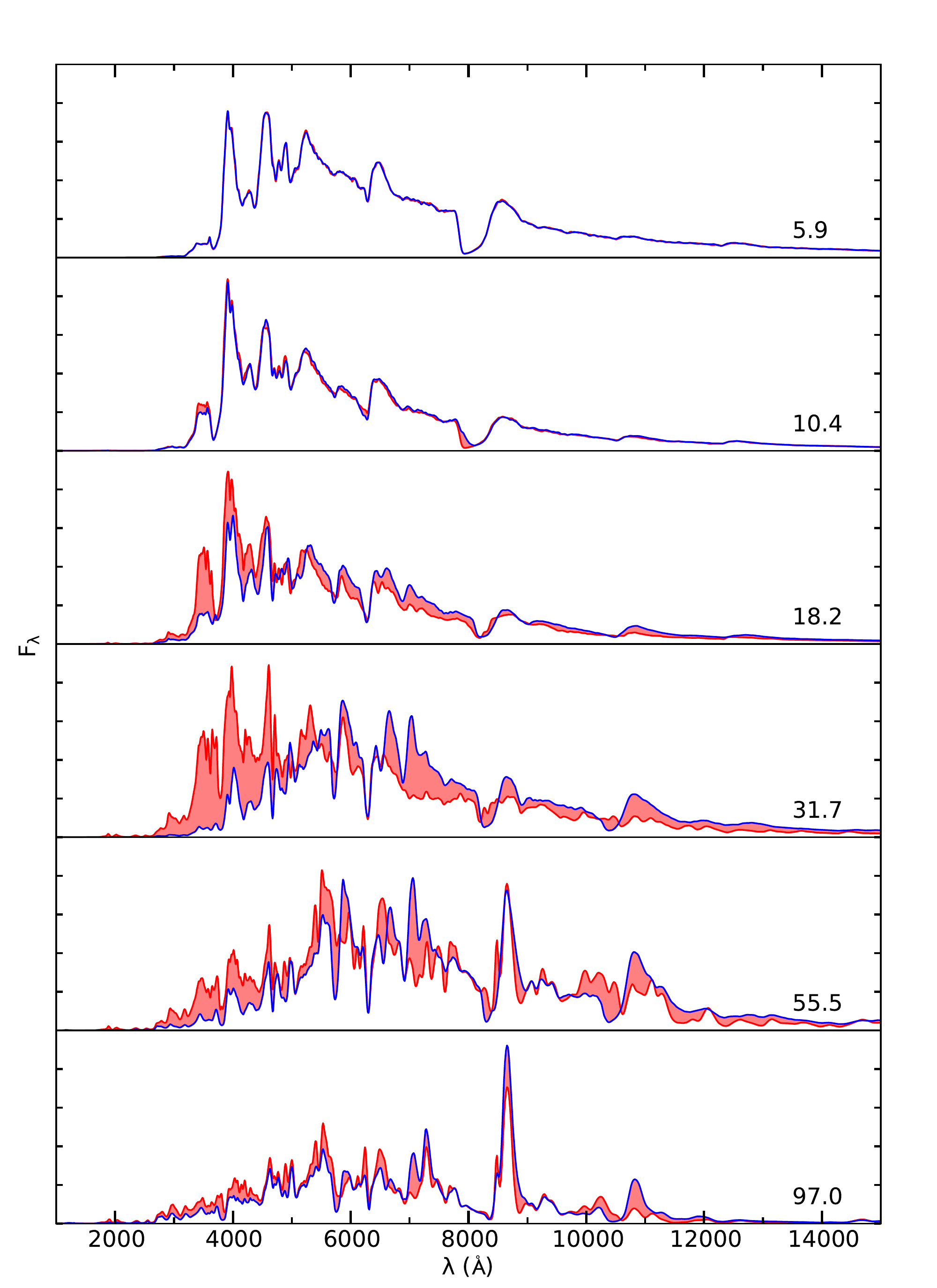}
\caption{Spectral evolution for model 12C calculated with (blue) and without (red) non-thermal ionization and excitation, where the difference has been highlighted in shaded red.}
\label{f_12C_nlte_comp_spec_evo}
\end{figure}

Figures \ref{f_12C_nlte_comp_lc_evo} and \ref{f_12C_nlte_comp_spec_evo} show the bolometric lightcurve and the spectral evolution of model 12C calculated with JEKYLL with and without non-thermal ionization and excitation. Before 10 days both the bolometric lightcurve and the spectral evolution are very similar, after which they start to differ in several aspects. This turning point coincides with the time when the radioactive energy deposition becomes important outside the photosphere (see Paper 2). The most striking difference in the spectral evolution is the absence of (strong) helium lines in the model without non-thermal processes. This well-known effect was pointed out already by \citet{Luc91}, and was later confirmed using CMFGEN by \citet{Des12}. Non-thermal excitation and ionization are essential to populate the excited levels of \ion{He}{i}, which is in turn required to produce the lines observed. As discussed by \citet{Luc91}, the population process is subtle, as ionization of \ion{He}{i} is amplified by photo-ionization from the excited levels, which proceeds at a rate far exceeding the non-thermal one.

\begin{figure}[tbp!]
\includegraphics[width=0.5\textwidth,angle=0]{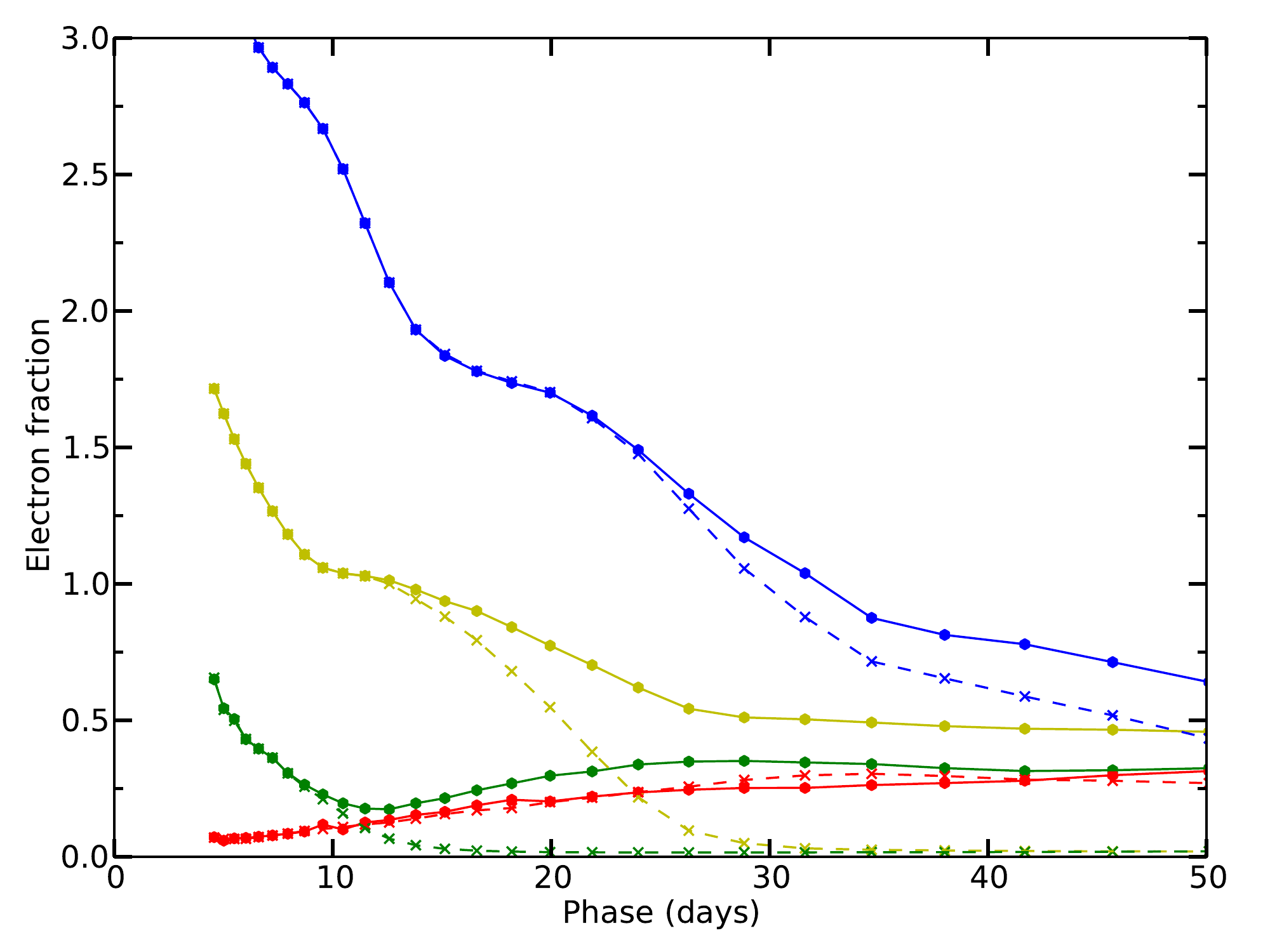}
\caption{Evolution of the electron fraction in the oxygen core (blue), inner/outer (yellow/green) helium envelope and the hydrogen envelope (red) for model 12C calculated with (circles and solid lines) and without (crosses and dashed lines) non-thermal ionization and excitation.}
\label{f_12C_nlte_comp_xe_evo}
\end{figure}

Less known is the quite strong effect on the bolometric lightcurve, where the diffusion peak of the model with non-thermal processes is considerably broader. The reason for this is the increased degree of ionization, and therefore the increased electron scattering opacity. This is illustrated by Fig.~\ref{f_12C_nlte_comp_xe_evo}, which shows the evolution of the electron fraction in the carbon-oxygen core and the helium and hydrogen envelopes. In the model with non-thermal processes, the electron fraction in the helium envelope drops much slower than in the model without. Due to the lower ionization potential, the effect is much less pronounced in the carbon-oxygen core and the hydrogen envelope. We note, that this means that the effect of non-thermal processes is probably largest in SNe dominated by species with high ionization potential (like He), so the case explored here might be somewhat extreme. In Fig.~\ref{f_12C_nlte_comp_lc_evo} we also show the bolometric lightcurve for the LTE version of model 12C used for the comparison with ARTIS in Sect.~\ref{s_comp_artis}. This model shows an even narrower bolometric lightcurve, which is related to an even lower degree of ionization. We note, that the atomic data for this model differs somewhat from that used for the NLTE models in this section, but this difference does not significantly affect the bolometric lightcurve.

\subsection{Convergence of the $\Lambda$-iterations}
\label{s_convergence}

\begin{figure}[tbp!]
\includegraphics[width=0.5\textwidth,angle=0]{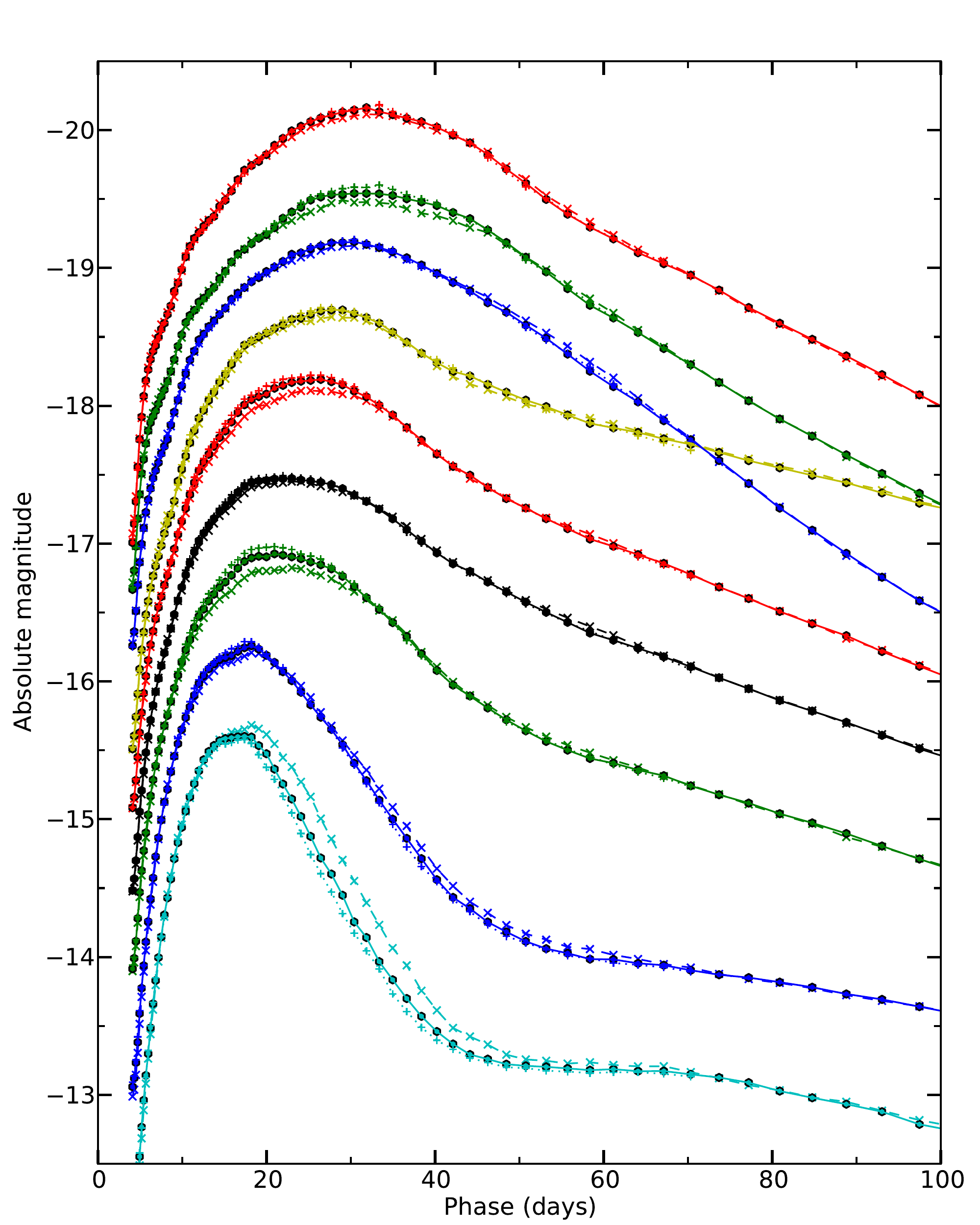}
\caption{Broad-band and bolometric lightcurve for model 12C as calculated with JEKYLL using two (dashed lines and crosses), four (solid lines and circles) and eight (dotted lines and pluses) $\Lambda$-iterations per time-step. Otherwise as described in Fig.~\ref{f_12C_lightcurve}.}
\label{f_12C_lambda_comp_lc_evo}
\end{figure}

\begin{figure}[tbp!]
\includegraphics[width=0.5\textwidth,angle=0]{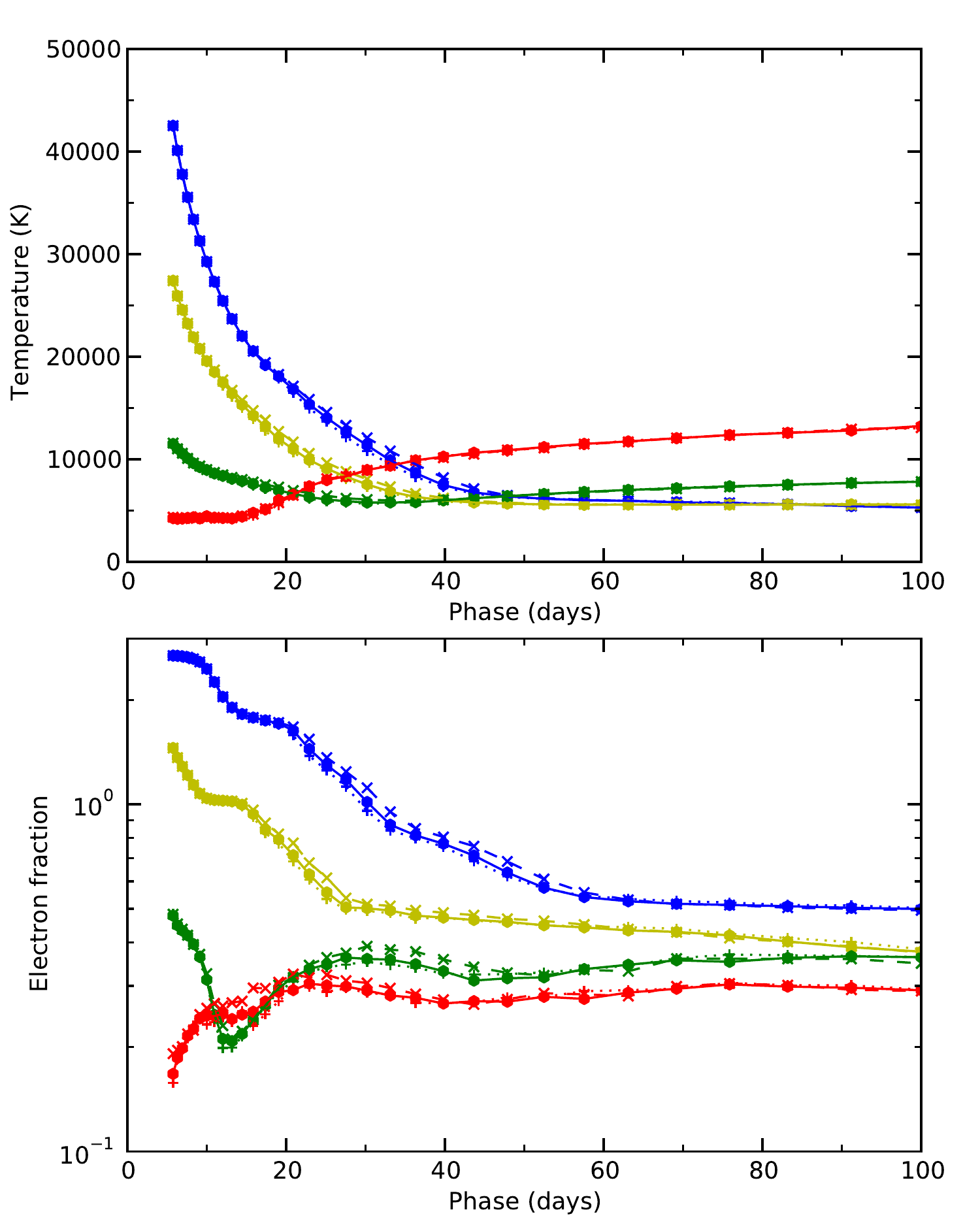}
\caption{Temperature (upper panel) and electron fraction (lower panel) for model 12C as calculated with JEKYLL using two (dashed lines and crosses), four (solid lines and circles) and eight (dotted lines and pluses) $\Lambda$-iterations per time-step. Otherwise as described in Fig.~\ref{f_12C_matter_evo}.}
\label{f_12C_lambda_comp_matter_evo}
\end{figure}

\begin{figure}[tbp!]
\includegraphics[width=0.5\textwidth,angle=0]{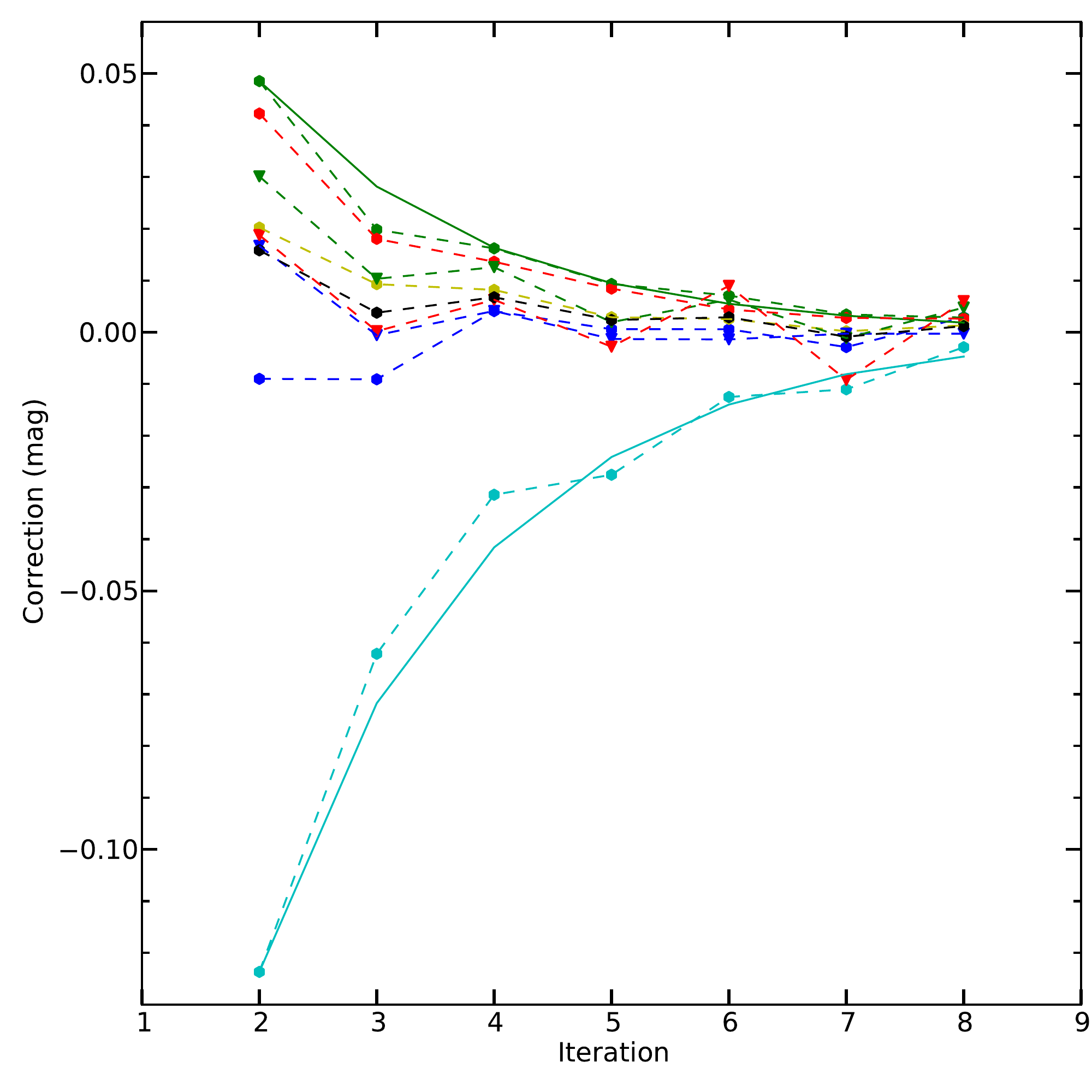}
\caption{Correction per $\Lambda$-iteration averaged between 15 and 40 days for the broad-band and bolometric lightcurves of model 12C. The colours used for the bands are the same as in Fig.~\ref{f_12C_lightcurve}, and the optical and NIR bands have been marked with circles and triangles, respectively. For the $U$- and $V$-bands we also show the curves of geometric convergence for a ratio of successive corrections of 0.58.}
\label{f_12C_lambda_comp_lc_conv}
\end{figure}

As mentioned in Sect.~\ref{s_method}, JEKYLL uses a fixed (but configurable) number of $\Lambda$-iterations. In time-dependent (radiative transfer) mode, this is the number of $\Lambda$-iterations per time-step, and corresponds to some (unknown) number of effective $\Lambda$-iterations depending on the length of the time-step and the rate at which the state is changing. The time-dependent NLTE run we present in this section uses a logarithmic time-step of 5 percent and four $\Lambda$-iterations per time-step. In Figs.~\ref{f_12C_lambda_comp_lc_evo} and \ref{f_12C_lambda_comp_matter_evo} we show the lightcurves and the temperature and electron fraction, respectively, for three such runs using two, four and eight $\Lambda$-iterations per time-step. We note, that to speed up the calculations we used coarser spatial sampling, slightly simplified atomic data and fewer packets than in the original model. As can be seen, convergence is fast, and more than four iterations per time-step does not make a significant difference. Even two iterations are good enough for most purposes, although there is a $\sim$25 percent difference in the $U$-band during the drop from the peak onto the tail. 

To further investigate the convergence, we plot the correction per $\Lambda$-iteration for the lightcurves in Fig.~\ref{f_12C_lambda_comp_lc_conv}. The corrections have been averaged between 15 and 40 days, where we have included the sign in the average to reduce the MC noise. As exemplified by the $U$ and $V$-band, which are the bands with the largest deviations, convergence appears to be geometric with a ratio of successive corrections close to 1/2. Given this ratio, the remaining error equals the last correction, which is less than one percent in all bands after eight $\Lambda$-iterations.

The comparisons to ARTIS in Sect.~\ref{s_comp_artis} behave in a similar way, but the shorter time-step of 1 percent, and possibly a faster convergence in the LTE case, make even a single-iteration run well converged, showing less discrepancy than the two-iterations run in Figure~\ref{f_12C_lambda_comp_lc_evo}. 

The comparisons to CMFGEN in Sect.~\ref{s_comp_cmfgen} also behave similarly. Using more than four $\Lambda$-iterations per time-step does not make a significant difference, and using four instead of two $\Lambda$-iterations only marginally changes the spectra. Furthermore, in this case we know that the $\Lambda$-iteration is converging to a solution that is (most likely) close to the correct one. Together with the nice convergence properties for model 12C, this is assuring, although further comparisons to CMFGEN would be interesting.

\subsection{Testing extensions of the Lucy method}

\subsubsection{The use of a diffusion solver}
\label{s_test_diffusion_solver}

The use of the diffusion solver in the inner region speeds up calculations in the early phase, and it is used in all simulations except the comparison with SUMO. It is therefore of interest to investigate how the diffusion solver, and the depth at which it is coupled to the MC radiative transfer solver, influence the solution. To achieve this, we have run model 12C with the diffusion solver coupled at a (Rosseland mean) continuum optical depth of 50, 100 and 200, using the same simplified set-up as described in Sect.~\ref{s_convergence}. At these coupling depths, the diffusion solver is only used until 21, 15 and 12 days, respectively, so in the last case most of the diffusion peak is actually calculated using the MC radiative transfer solver alone. 

The results show no significant differences in the spectra and only small differences in the matter quantities, which justifies the use of the diffusion solver, at least at an optical depth of 50 or more. It is also interesting to return to the comparison with CMFGEN in Fig.~\ref{f_cmfgen_comp_matter_evo}, where we have marked the border between the diffusion solver and the MC radiative transfer solver. As we can here compare to a solution that is (most likely) close to the correct one, it is evident that the use of a coupled diffusion and MC radiative transfer solver works well.

\subsubsection{The packet sampling control}
\label{s_test_packet_control}

\begin{figure}[tbp!]
\includegraphics[width=0.5\textwidth,angle=0]{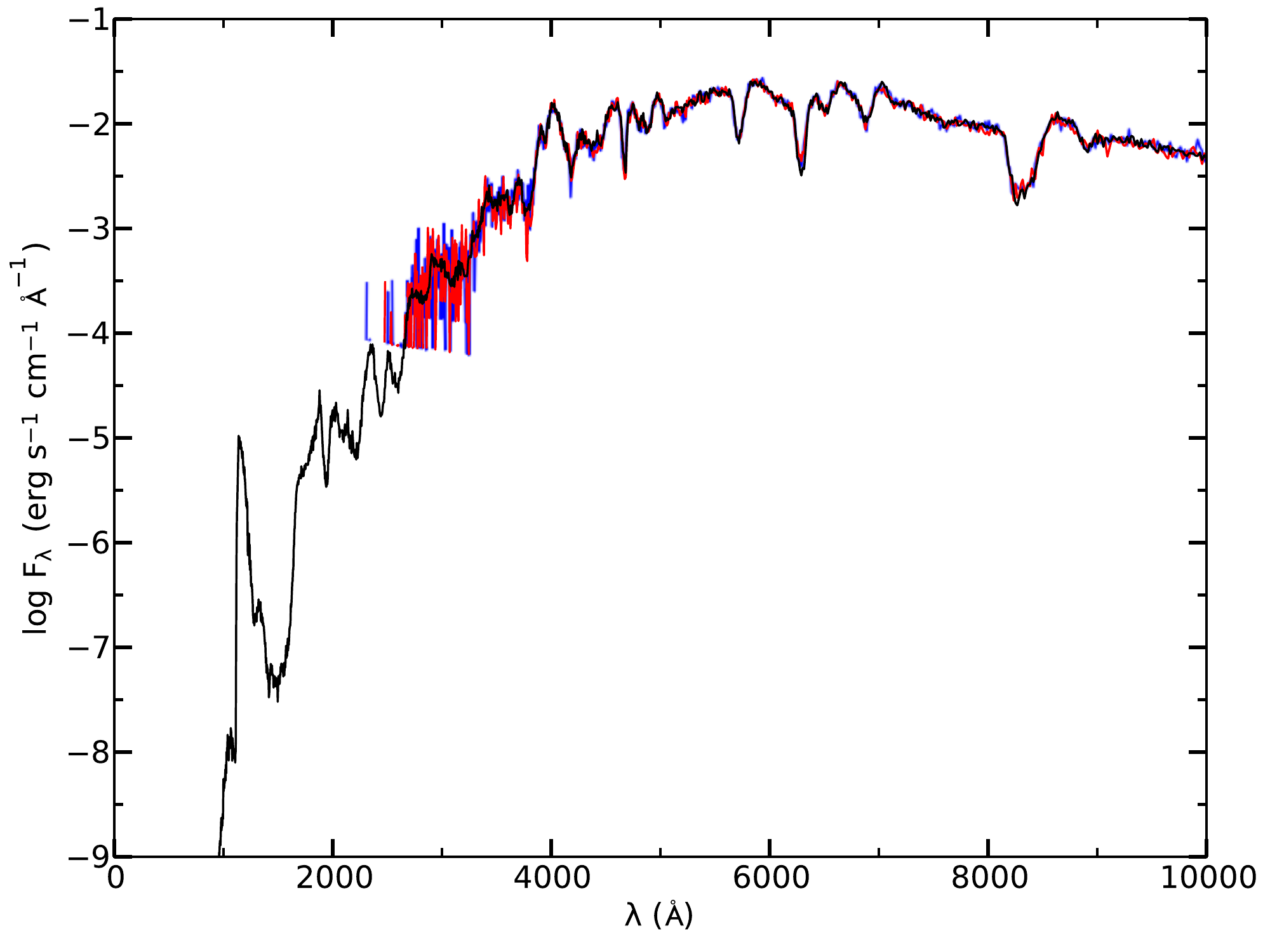}
\caption{Spectra for model 12C at 34.7 days with (black) and without (red) packet sampling control activated, as well as with (red) and without (blue) the Markov-Chain solution activated.}
\label{f_12C_boost_comp_spec}
\end{figure}

The method for packet sampling control was introduced in Sect.~\ref{s_packet_control}, and is used to decrease the noise in the radiation field estimators. As a test, we have re-run the MC radiative transfer for model 12C with and without packet sampling control activated. In doing this we have loaded the matter state from the original model 12C run and kept it fixed. Figure~\ref{f_12C_boost_comp_spec} shows the spectrum for model 12C at 34.7 days with and without packet sampling control activated. The packet sampling control was configured to maintain a SNR of 3 percent between the Lyman break and 25000 \AA~with a minimum boost factor of 1 and and a maximum boost factor of 10$^{12}$. As seen in Fig.~\ref{f_12C_boost_comp_spec}, the two spectra agree well as long as the SNR is good in both, which shows that the method reproduces the correct radiation field. In addition, in the region bluewards $\sim$3000 \AA, a SNR of $\sim$3 percent is maintained all the way to the Lyman break in the model with packet sampling control, but in the model without there are no MC packets at all below $\sim$2500 \AA. 

In the model with packet sampling control, the average boost of the number of packets is $\sim$4, whereas the boost factors in the blue region approach $\sim$10$^{9}$ near the Lyman break. Without packet sampling control the same SNR in the blue region could only have been achieved by increasing the total number of packets by this huge factor. Finally, we repeated the test, but decreased the number of packets injected into the calculation with a factor of ten. The resulting spectrum is indistinguishable from that from the original test, the only difference being the ten times higher boost factors needed to achieve the required SNR.

\subsubsection{The Markov-Chain solution}
\label{s_test_markov_chain}

The Markov-Chain solution to the MC state-machine was introduced in Sect.~\ref{s_markov_chains} as a way to sample the emission frequency more efficiently. As a test, we have re-run the MC radiative transfer for model 12C with and without the Markov-Chain solution activated. As in Sect.~\ref{s_test_packet_control}, we have loaded the matter state from the original model 12C run and kept it fixed. As for most simulations in this paper, we have used the split state-machine approach (see Sect.~\ref{s_markov_chains} and Appendix~\ref{a_markov_chains}) when calculating the Markov-chain solution, as this speeds up the calculation by a large factor.

Figure~\ref{f_12C_boost_comp_spec} shows the spectrum for model 12C at 34.7 days with and without the Markov-Chain solution activated. The spectra agree well, which shows that the Markov-Chain solution produces the same radiation field as the original state-machine. The speed-up of the MC radiative transfer when using the Markov-Chain solution is a factor of $\sim$50 at 1 day, increases to $\sim$150 at 4 days and then declines and levels out at $\sim$10. The overhead from calculating the Markov-chain solution (using the split state-machine approach) is small at early times, but increase with time, and at 40 days it decrease the effective speed-up with $\sim$30 percent. The number of packets as well as the depth at which the diffusion solver is coupled affects the measured speed-up, but the case investigated here should be fairly representative. Clearly the method is essential at early times, and otherwise most helpful in reducing the computational effort.

\section{Conclusions}
\label{s_conclusions}

We have presented and described JEKYLL, a new code for modelling of SN spectra and lightcurves. The code assumes homologous expansion, spherical symmetry and steady state for the matter, but is otherwise capable of solving for the time-evolution of the matter and the radiation field in full NLTE. The method used is an extension of the MC based Lucy method, here tested for the first time in its time-dependent NLTE version. In particular, it includes a detailed treatment of non-thermal excitation and ionization, where the rates are calculated as described in \citetalias{Koz92}. Another important feature is a method to account for the macroscopic mixing that occurs in the explosion, which was previously introduced in \citetalias{Jer11}. We have also described how to speed up the calculation by using a diffusion solver in the inner region, and by using Markov-chains to sample the packet frequency more efficiently. In addition, we have introduced a novel method to control the sampling of the radiation field, which is used to reduce the noise in the radiation field estimators.

We have presented comparisons with the ARTIS, SUMO and CMFGEN codes. The ARTIS and SUMO codes are similar to JEKYLL in some, but not all, aspects, and the comparisons provide tests of the time-dependent MC radiative transfer and the steady-state NLTE capabilities, respectively. The CMFGEN code is similar in terms of physics, but uses a different method, where the coupled system of differential equations for the matter and the radiation field is solved by a linearization technique. This comparison, which was done with a somewhat simplified ejecta model, provides a test of the time-dependent NLTE capabilities of JEKYLL. All comparisons show a good agreement in the observed quantities, as well as the state variables, and together they provide a thorough test of the JEKYLL code. In particular, the comparison with CMFGEN shows that the MC based Lucy method, where the de-coupled radiation-matter problem is solved through $\Lambda$-iteration, does indeed converge in the time-dependent NLTE case. This has previously only been shown for the steady-state NLTE case in a hydrogen SN atmosphere \citepalias{Luc03}.

Finally, we have presented an example of the time-dependent NLTE capabilities of JEKYLL using a realistic ejecta model for a Type IIb SN. This model belongs to the set of Type IIb models presented by \citetalias{Jer15}, which will be explored in more detail and compared to observations in Paper 2. Based on the model we find strong effects of NLTE even on the bolometric lightcurve. Although the case explored might be somewhat extreme, this casts some doubts on LTE-based modelling of the bolometric lightcurve commonly used in the literature. For example non-thermal ionization turns out to have a strong effect on the ionization level in the helium envelope, which introduces a coupling between the mixing of the radioactive $^{56}$Ni and the diffusion time not accounted for in LTE-based models. 

We also use the model to test our most important (technical) extensions of the Lucy method. Among others, these tests show that the Markov-chain solution to the MC state-machine may speed up the calculations in the early phase by large factors, and that the method for packet sampling control may improve the SNR in noisy wavelength regions by huge factors at a modest computational cost.

\section{Acknowledgements}

Our work makes ample use of techniques that were developed by Leon Lucy, who sadly passed away earlier this year. We want to thank him for his pioneering contributions to MC radiative transfer (as well as to many other fields of astrophysics).

We also thank John Hillier for help and discussion on CMFGEN and the comparison with JEKYLL in Sect.~\ref{s_comp_cmfgen}.

This research was supported by the Swedish Research Council and the Swedish National Space Board, and the computations were performed on resources provided by the Swedish National Infrastructure for Computing (SNIC) at Parallelldatorcentrum (PDC).

\appendix

\section{Configuration and atomic data}
\label{a_config_atomic_data}

\subsection{Comparison to ARTIS} 
\label{a_conf_data_comp_artis}

To synchronize JEKYLL with ARTIS, both codes were configured to use a LTE solution for the matter based on $T_\mathrm{J}$, the temperature associated with the pure black-body radiation field model (see Sect.~\ref{s_cell} and \citetalias{Kro09}), and the MC radiative transfer solver used by JEKYLL was configured to include radiative and collisional bound-bound and bound-free processes as well as free-free processes. In addition, ARTIS was configured to use its grey approximation (see \citetalias{Kro09}) before $\sim$6 days and below an optical depth of 100, and JEKYLL was configured to use the diffusion solver below an optical depth of 50. 

The atomic data used by ARTIS is described in \citetalias{Kro09}, but was restricted to the first four ionization stages, using the fourth as closure. As all of the atomic data is stored in data-files in a well-defined format, it was quite straight-forward to automatically convert it to the JEKYLL atomic data format, and it should be fully synchronized. ARTIS and JEKYLL were both configured to use a logarithmic time-step of 1 percent and single $\Lambda$-iteration per time-step, which is the standard procedure in ARTIS. As discussed in Sect~\ref{s_convergence}, due to the short time-step, these runs are still well converged.

\subsection{Comparison to SUMO} 
\label{a_conf_data_comp_sumo}

As much as possible, we have synchronized the configuration and the atomic data used by JEKYLL with that used for the modelling in \citetalias{Jer15}. To achieve this, JEKYLL was configured to run in steady-state mode, and to use a full NLTE solution including the following; radiative bound-bound, bound-free and free-free processes, collisional bound-bound processes, non-thermal excitation, ionization and heating, as well as charge-transfer and two-photon processes. JEKYLL was also configured to use a recombination correction in a manner similar to SUMO (see \citetalias{Jer11}), in which case detailed balance was not enforced.

The atomic data used for the modelling in \citetalias{Jer15} is described in \citetalias{Jer11} and \citet{Jer12}. In the case it was stored in data-files in a well-defined format, as for for example energy levels and spontaneous emission rates, it was automatically converted to the JEKYLL atomic data format, and otherwise it was added manually to the JEKYLL atomic data files based on the descriptions in \citetalias{Jer11} and \citet{Jer12}. Although not complete, the synchronization of the atomic data and the methods should be good enough for a meaningful comparison.

\subsection{Comparison to CMFGEN} 
\label{a_conf_data_comp_cmfgen}

To synchronize JEKYLL with CMFGEN, JEKYLL was configured run in time-dependent (radiative transfer) mode, and to use a full NLTE solution including the following; radiative bound-bound, bound-free and free-free processes, as well as collisional bound-bound and bound-free processes. JEKYLL was also configured to use the time-dependent NLTE rate equations and to use the diffusion solver below an optical depth of 100. In addition, to assure good sampling of the radiation field, packet control (see Sect.~\ref{s_packet_control}) was turned on and the generalized on-the-spot approximation (see Sect.~\ref{s_cell}) was used bluewards the Lyman break.

The atomic data for the simplified composition of hydrogen, helium, oxygen and calcium were automatically converted from the well-defined format of CMFGEN to that of JEKYLL, and should therefore be fully synchronized. CMFGEN and JEKYLL were both configured to use a logarithmic time-step of 2.5 percent, and JEKYLL was configured to use 4 $\Lambda$-iterations per time-step. As discussed in Sect.~\ref{s_convergence}, this gives a well converged solution.

\subsection{Application to type IIb SNe} 
\label{a_conf_data_app}

JEKYLL was configured to run in time-dependent (radiative transfer) mode, and to use a full NLTE solution including the following; radiative bound-bound, bound-free and free-free processes, collisional bound-bound and bound-free processes, non-thermal excitation, ionization and heating, as well as two-photon processes. JEKYLL was also configured to use the diffusion solver below an optical depth of 50, and to use a recombination correction while still enforcing detailed balance. In addition, to assure good sampling of the radiation field, packet control (see Sect.~\ref{s_packet_control}) was turned on and the generalized on-the-spot approximation (see Sect.~\ref{s_cell}) was used bluewards the Lyman break. The logarithmic time-step was set to 5 percent and the number of $\Lambda$-iterations per time-step was set to 4. As discussed in Sect.~\ref{s_convergence}, this gives a well converged solution. 

The atomic data used is the same as for the comparison with SUMO (Sect.~\ref{a_conf_data_comp_sumo}), but with the following modifications. The highest ionization stage was increased to VI for all species, and the stage III ions were updated to include at least 50 levels for elements lighter than Sc, and at least 200 levels for heavier elements, using online data provided by NIST\footnote{www.nist.gov} and R. Kurucz\footnote{http://www.cfa.harvard.edu/amp/ampdata/kurucz23/sekur.html}. Total recombination rates for the stage III ions were adopted from the online table provided by S. Nahar\footnote{http://www.astronomy.ohio-state.edu/\texttt{\char`\~}nahar/\texttt{\char`\_}naharradiativeatomicdata/} whenever available, and otherwise from \citet{Shu82}. For ionization stages IV to VI we only included the ground-state multiplets, adopted the photo-ionization cross-section by \citet{Ver95} and \citet{Ver96} and assumed the populations to be in LTE with respect to stage IV.

\section{Splitting the MC state-machine}
\label{a_markov_chains}

To split the MC state-machine and the corresponding Markov-chain model into its base- and sub-machine parts (Sect.~\ref{s_packet_interaction}), we proceed as follows. First, consider the base-machine, where the states correspond to the macro-atoms and the thermal pool. The probabilities for internal transitions and de-activation from the thermal pool are given by the cooling rates. More specifically, the probability for an internal transition from the thermal pool (labelled $T$) to macro-atom state $I$ is given by 
\begin{equation}
P^{\mathrm{I}}_{T \rightarrow I}=\sum\limits_{i} P^{\mathrm{A,C}}_{I,i}
\end{equation}
where $P^{\mathrm{A,C}}_{I,i}$ is the probability for collisional activation of macro-atom sub-machine $I$ in state $i$, in turn given by the probability for collisional cooling through an upward transition to state $i$ of macro-atom sub-machine $I$. The probability for de-activation from the thermal pool, $P^{\mathrm{D}}_{T}$, is just the probability for radiative cooling. In the case that the base-machine is activated in its thermal pool state (which is the case we are interested in), the probabilities for internal transitions and de-activations from the macro-atom states are given by the cooling rates and the $\mathbf{SR}$ matrices (Sect.~\ref{s_markov_chains}) for the macro-atom sub-machines. More specifically, the probability for an internal transition from macro-atom state $I$ to the thermal pool is
\begin{equation}
P^{\mathrm{I}}_{I \rightarrow T}=\sum\limits_{i,j} P^{\mathrm{A,C}}_{I,i} (\mathbf{SR})^{\mathrm{C}}_{I,i,j}
\end{equation}
where $\mathbf{SR}^{\mathrm{C}}_{I}$ is the $\mathbf{SR}$ matrix for macro-atom sub-machine $I$ with respect to \textit{collisional} de-activation. Similarly, the probability for de-activation from macro-atom state  $I$ is given by
\begin{equation}
P^{\mathrm{D}}_{I}=\sum\limits_{i,j} P^{\mathrm{A,C}}_{I,i} (\mathbf{SR})^{\mathrm{R}}_{I,i,j}
\end{equation}
where $\mathbf{SR}^{\mathrm{R}}_{I}$ is the $\mathbf{SR}$ matrix for macro-atom sub-machine $I$ with respect to \textit{radiative} de-activation. From the probabilities for internal transitions and de-activations we can calculate the base-machine $\mathbf{SR}$ matrix as described in Sect.~\ref{s_markov_chains}. We note, that this is done for the case when the base-machine is activated in its thermal pool state, which is all we need. 

The procedure is now as follows. If the base-machine is activated in a macro-atom state, a macro-atom sub-machine is activated by a radiative transition. The de-activating transition is then drawn based on the $\mathbf{SR}$ matrix and the de-activation probabilities for this macro-atom sub-machine, as described in Sect.~\ref{s_markov_chains}. Radiative de-activation corresponds to de-activation of the base-machine, whereas collisional de-activation corresponds to an internal base-machine transition to the thermal pool. In the latter case, or if the base-machine was activated in its thermal pool state, the state from which the base-machine de-activates is drawn based on its $\mathbf{SR}$ matrix, as described in Sect.~\ref{s_markov_chains}. If it de-activates from the thermal pool, an emission process is drawn in proportion to the radiative cooling rates. If it de-activates from a macro-atom state, the de-activating transition is drawn based on the $\mathbf{SR}^{\mathrm{R}}$ matrix and the \textit{radiative} de-activation probabilities for the macro-atom sub-machine, as described in Sect.~\ref{s_markov_chains}. The split state-machine approach is more involved and comes at the expense of (two) more draws, but at a greatly reduced cost to calculate and store the $\mathbf{SR}$ matrices, and has been used for most of the simulations in this paper.

\section{Adjusting the MC emission rates}
\label{a_packet_control}

As explained in Sect.~\ref{s_packet_control}, control of the number of packets is achieved by adjusting their size (i.e.~the amount of energy they carry, see Sect.~\ref{s_packet_control}) as a function of frequency\footnote{The variation of the packet size in space and time is ignored in this section.}. This may be expressed through a boost factor $B(\nu)$, in terms of which the packet size is given by $E_\mathrm{P}(\nu)=E_{\mathrm{P},0}/B(\nu)$, where $E_{\mathrm{P},0}$ is some reference size. If $B(\nu)>1$, this corresponds to a boost of the number of packets, and if $B(\nu)<1$ it corresponds to a reduction. As discussed in Sect.~\ref{s_packet_control}, the values for the boost factors\footnote{The discussion in Sect.~\ref{s_packet_control} is in terms of packet size, but that makes no difference.} used in JEKYLL are determined by an adaptive algorithm, in which they are adjusted once per $\Lambda$-iteration to achieve a pre-configured target SNR.

As explained in Sect.~\ref{s_packet_control}, to conserve energy, the emissivity in the destination region (in terms of packets) has to be adjusted with $F$, the ratio of the packet sizes in the source and destination regions. In terms of the boost factors in these regions, $F$ may also be written as a boost ratio, that is $F(\nu_A,\nu_E)=B(\nu_E)/B(\nu_A)$, where $\nu_A$ and $\nu_E$ are the absorption and emission frequencies of the packet. As is also explained in Sect.~\ref{s_packet_control}, each emission event in the destination region is triggered by a fictitious absorption event in the source region. For a given interaction process, these fictitious absorption events correspond to an fictitious opacity $\kappa_{\mathrm{F}}$, which gives the interaction rate required to obtain the adjusted emissivities.

Beginning with absorption, a packet takes one of the possible paths through the MC state-machine, and is eventually emitted. To adjust the rates of packets flowing through the MC state-machine, we need to proceed in the reverse order. First, based on the boost factor ($B$), boost factors for the emissivities ($B^\mathrm{E}$) are calculated and adjusted accordingly. Second, based on the boost factors for the emissivities, boost factors for the MC state-machine probabilities ($B^\mathrm{D}$, $B^\mathrm{A}$ and $B^\mathrm{C}$) are calculated and adjusted accordingly. Finally, based on the boost factors for the MC state-machine probabilities, boost factors for the opacities ($B^\mathrm{O}$) are calculated. For a specific packet with frequency $\nu$, this gives the corresponding boost ratios $F^\mathrm{O}(\nu)=B^\mathrm{O}/B(\nu)$, which in turn give the fictitious opacities $\kappa_{\mathrm{F}}(\nu)=F^{\mathrm{O}}(\nu)~\kappa(\nu)$. However, as absorption has to proceed at the original rate, the fictitious opacity to use is not $\kappa_\mathrm{F}$ but $max(\kappa_{\mathrm{F}},\kappa)=max(F^{\mathrm{O}},1)~\kappa$. Using this fictitious opacity, packets are then selected for emission, absorption or both according to the rules described in Sect.~\ref{s_packet_control}.

We note, that in this appendix we follow the convention to number energy levels with respect to the macro-atoms instead of the ions. We also note, that the boost factors and the adjustment of the macro-atom probabilities in Appendix~\ref{a_pc_macro_atom} apply to the complete MC state-machine, and have to be adjusted in case the split MC state-machine approach (see Appendix.~\ref{a_markov_chains}) is used.

\subsection{Emissivities}
\label{a_pc_emissivities}

The boost factor for emission as a function of frequency is just $B(\nu)$, and from this the boost factors for bound-bound, bound-free and free-free emission are calculated as
\begin{equation}
\begin{split}
&B^\mathrm{E,BB}_{I,i \rightarrow j}=B(\nu_{I,i \rightarrow j})\\
&B^\mathrm{E,BF}_{I,i \rightarrow j}=\int B(\nu)~f^\mathrm{BF}_{I,i \rightarrow j}(\nu)~d\nu\\
&B^\mathrm{E,FF}=\int B(\nu)~f^\mathrm{FF}(\nu)~d\nu
\end{split}
\end{equation}
where $f^\mathrm{BF}_{I,i \rightarrow j}(\nu)$ and $f^\mathrm{FF}(\nu)$ are the distribution functions for bound-free and free-free emission, respectively. The distribution functions for bound-free and free-free emission are then adjusted as
\begin{equation}
\begin{split}
&f'^{\mathrm{BF}}_{I,i \rightarrow j}(\nu)=B(\nu)~f^\mathrm{BF}_{I,i \rightarrow j}(\nu)~/B^\mathrm{E,BF}_{I,i \rightarrow j}\\
&f'^{\mathrm{FF}}(\nu)=B(\nu)~f^\mathrm{FF}(\nu)~/B^\mathrm{E,FF}
\end{split}
\end{equation}

\subsection{Macro-atom probabilities}
\label{a_pc_macro_atom}

The boost factor for de-activation from state $i$ of macro-atom $I$ is calculated as
\begin{equation}
B^{\mathrm{D}}_{I,i}=\sum\limits_{j} B^\mathrm{E}_{I,i \rightarrow j} P^{\mathrm{D}}_{I,i \rightarrow j}
\end{equation}
where $P^{\mathrm{D}}_{I,i \rightarrow j}$ is the probability for a de-activating (bound-bound or bound-free) transitions from state $i$ to state $j$ of macro-atom $I$. The probability for de-activating transitions is then adjusted as 
\begin{equation}
P'^{\mathrm{D}}_{I,i \rightarrow j}=B^\mathrm{E}_{I,i \rightarrow j} P^{\mathrm{D}}_{I,i \rightarrow j} / B^{\mathrm{D}}_{I,i}
\end{equation}
The boost factor for activation of macro-atom $I$ in state $j$ is calculated as
\begin{equation}
B^{\mathrm{A}}_{I,i}=\sum\limits_{J,j} B^{\mathrm{D}}_{J,j}~(\mathbf{SR})_{(I,i),(J,j)}+B^{\mathrm{C,R}}~(\mathbf{SR})_{(I,i),T}
\end{equation}
where the $\mathbf{SR}$ matrix is obtained from a Markov-chain solution to the MC state-machine (see Sect.\ref{s_markov_chains}), and $B^{\mathrm{C,R}}$ is the boost factor for radiative cooling (see Appendix~\ref{a_pc_thermal_pool}). The indices of $\mathbf{SR}$ are labelled $(N,n)$ if they correspond to state $n$ of macro-atom $N$ or $T$ if they correspond to the thermal pool. The $\mathbf{SR}$ matrix is then adjusted as 
\begin{equation}
\begin{split}
&(\mathbf{SR})'_{(I,i),(J,j)}=B^{\mathrm{D}}_{J,j}~(\mathbf{SR})_{(I,i),(J,j)}~/B^{\mathrm{A}}_{I,i}\\
&(\mathbf{SR})'_{(I,i),T}=B^{\mathrm{C,R}}~(\mathbf{SR})_{(I,i),T}~/B^{\mathrm{A}}_{I,i}
\end{split}
\end{equation}

\subsection{Thermal pool probabilities}
\label{a_pc_thermal_pool}

The boost factor for collisional cooling is calculated as
\begin{equation}
B^{\mathrm{C,C}}=\sum\limits_{I,i,j} B^{\mathrm{A}}_{I,j} P^\mathrm{C,C}_{I,i \rightarrow j}
\end{equation}
where $P^\mathrm{C,C}_{I,i \rightarrow j}$ is the probability for collisional cooling through a transition from state $i$ to state $j$ of macro-atom $I$. This probability is then adjusted as 
\begin{equation}
P'^{\mathrm{C,C}}_{I,i \rightarrow j}=B^{\mathrm{A}}_{I,j} P^\mathrm{C,C}_{I,i \rightarrow j}~/B^{\mathrm{C,C}}
\end{equation}
The boost factor for radiative cooling is calculated as 
\begin{equation}
B^{\mathrm{C,R}}=\sum\limits_{I,i,j} B^{\mathrm{E,BF}}_{I,i \rightarrow j} P^\mathrm{C,BF}_{I,i \rightarrow j}+B^{\mathrm{E,FF}} P^\mathrm{C,FF}
\end{equation}
where $P^\mathrm{C,BF}_{I,i \rightarrow j}$ is the probability for radiative cooling through a bound-free transition from state $i$ to state$j$ of macro-atom $I$ and $P^\mathrm{C,FF}$ is the probability for radiative cooling through free-free emission. These probabilities are then adjusted as 
\begin{equation}
\begin{split}
&P'^{\mathrm{C,BF}}_{I,i \rightarrow j}=B^{\mathrm{E,BF}}_{I,i \rightarrow j} P^\mathrm{C,BF}_{I,i \rightarrow j}~/B^{\mathrm{C,R}}\\
&P'^{\mathrm{C,FF}}=B^{\mathrm{E,FF}} P^\mathrm{C,FF}~/B^{\mathrm{C,R}}
\end{split}
\end{equation}
The boost factor for (the total) cooling is calculated as 
\begin{equation}
B^{\mathrm{C}}=B^{\mathrm{C,C}} P^{\mathrm{C,C}}+B^{\mathrm{C,R}} P^{\mathrm{C,R}}
\end{equation}
where $P^{\mathrm{C,C}}$ and $P^{\mathrm{C,R}}$ are the probabilities for collisional and radiative cooling, respectively. The probabilities for collisional and radiative cooling are then adjusted as 
\begin{equation}
\begin{split}
&P'^{\mathrm{C,C}}=B^{\mathrm{C,C}} P^\mathrm{C,C}~/B^{\mathrm{C}}\\
&P'^{\mathrm{C,R}}=B^{\mathrm{C,R}} P^\mathrm{C,R}~/B^{\mathrm{C}}
\end{split}
\end{equation}

\subsection{R-packet opacities}
\label{a_pc_r_opacities}

The boost factor for absorption through a bound-bound transition from state $i$ to state $j$ of macro-atom $I$ is calculated as 
\begin{equation}
B^{\mathrm{O,BB}}_{I,i \rightarrow j}=B^{\mathrm{A}}_{I,j}
\end{equation}
and the fictitious Sobolev optical depth for bound-bound transitions is then calculated as
\begin{equation}
\tau^{\mathrm{F}}_{I,i \rightarrow j}=max(F^{\mathrm{O,BB}}_{I,i \rightarrow j},1)~\tau_{I,i \rightarrow j}
\end{equation}
The boost factor for absorption through a bound-free transition from state $i$ to state $j$ of macro-atom $I$ is calculated as 
\begin{equation}
B^{\mathrm{O,BF}}_{I,i \rightarrow j}=B^{\mathrm{A}}_{I,j} P^{\mathrm{I}}_{I,i \rightarrow j}+B^{\mathrm{C}} P^{\mathrm{H}}_{I,i \rightarrow j}
\end{equation}
where $P^{\mathrm{I}}_{I,i \rightarrow j}$ and $P^{\mathrm{H}}_{I,i \rightarrow j}$ are the probabilities that the bound-free transition results in ionization and heating, respectively. The probabilities for ionization and heating is then adjusted as 
\begin{equation}
\begin{split}
&P'^{\mathrm{I}}_{I,i \rightarrow j}=B^{\mathrm{A}}_{I,j} P^{\mathrm{I}}_{I,i \rightarrow j}~/B^{\mathrm{O,BF}}_{I,i \rightarrow j}\\
&P'^{\mathrm{H}}_{I,i \rightarrow j}=B^{\mathrm{C}} P^{\mathrm{H}}_{I,i \rightarrow j}~/B^{\mathrm{O,BF}}_{I,i \rightarrow j}
\end{split}
\end{equation}
The boost factor for bound-free absorption is calculated as 
\begin{equation}
B^{\mathrm{O,BF}}(\nu)=\sum\limits_{I,i,j} B^{\mathrm{O,BF}}_{I,i \rightarrow j} \kappa_{I,i \rightarrow j}(\nu) / \sum\limits_{I,i,j} \kappa_{I,i \rightarrow j}(\nu)
\end{equation}
and the fictitious opacity for bound-free absorption is then calculated as 
\begin{equation}
\kappa_{\mathrm{F}}^{\mathrm{BF}}(\nu)=max(F^{\mathrm{O,BF}}(\nu),1)~\kappa^{\mathrm{BF}}(\nu)
\end{equation}
The boost factor for free-free absorption is calculated as
\begin{equation}
B^{\mathrm{O,FF}}=B^{\mathrm{C}}
\end{equation}
and the fictitious opacity for free-free absorption is then calculated as 
\begin{equation}
\kappa_{\mathrm{F}}^{\mathrm{FF}}(\nu)=max(F^{\mathrm{O,FF}},1)~\kappa^{\mathrm{FF}}(\nu)
\end{equation}

\subsection{$\gamma$-packet opacities}
\label{a_pc_g_opacities}

The boost factor for absorption of a $\gamma$-packet is calculated as
\begin{equation}
B^{\mathrm{O,G}}=B^{\mathrm{C}} P^{\mathrm{NT,H}}+\sum\limits_{I,i,j} B^{\mathrm{A,BB}}_{I,i \rightarrow j} P^{\mathrm{NT,E}}_{I,i \rightarrow j}+\sum\limits_{I,i,j} B^{\mathrm{A,BF}}_{I,i \rightarrow j} P^{\mathrm{NT,I}}_{I,i \rightarrow j}
\end{equation}
where $P^{\mathrm{NT,H}}$ and $P^{\mathrm{NT,E/I}}_{I,i \rightarrow j}$ are the probabilities for non-thermal heating and non-thermal excitation/ionization through a (bound-bound/bound-free) transition from state $i$ to state $j$ of macro-atom $I$. The probabilities for non-thermal heating, excitation and ionization are then adjusted as 
\begin{equation}
\begin{split}
&P'^{\mathrm{NT,H}}=B^{\mathrm{C}} P^{\mathrm{NT,H}}~/B^{\mathrm{O,G}}\\
&P'^{\mathrm{NT,E}}_{I,i \rightarrow j}=B^{\mathrm{A,BB}}_{I,j} P^{\mathrm{NT,E}}_{I,i \rightarrow j}~/B^{\mathrm{O,G}}\\
&P'^{\mathrm{NT,I}}_{I,i \rightarrow j}=B^{\mathrm{A,BF}}_{I,j} P^{\mathrm{NT,I}}_{I,i \rightarrow j}~/B^{\mathrm{O,G}}
\end{split}
\end{equation}
and from the effective opacity used for the g-packet (which is different for different decays and decay products), the corresponding fictitious opacity is calculated as
\begin{equation}
\kappa_{\mathrm{F}}^{\mathrm{G}}(\nu)=max(F^{\mathrm{O,G}},1)~\kappa^{\mathrm{G}}(\nu)
\end{equation}

\bibliographystyle{aa}
\bibliography{jekyll}

\begin{thebibliography}{51}
\expandafter\ifx\csname natexlab\endcsname\relax\def\natexlab#1{#1}\fi

\bibitem[{{Abbott} \& {Lucy}(1985)}]{Abb85}
{Abbott}, D.~C. \& {Lucy}, L.~B. 1985, \apj, 288, 679

\bibitem[{{Avrett} \& {Loeser}(1988)}]{Avr88}
{Avrett}, E.~H. \& {Loeser}, R. 1988, \apj, 331, 211

\bibitem[{{Bersten} {et~al.}(2012){Bersten}, {Benvenuto}, {Nomoto}, {Ergon},
  {Folatelli}, {Sollerman}, {Benetti}, {Botticella}, {Fraser}, {Kotak},
  {Maeda}, {Ochner}, \& {Tomasella}}]{Ber12}
{Bersten}, M.~C., {Benvenuto}, O.~G., {Nomoto}, K., {et~al.} 2012, \apj, 757,
  31

\bibitem[{{Cannon}(1973{\natexlab{a}})}]{Can73b}
{Cannon}, C.~J. 1973{\natexlab{a}}, \jqsrt, 13, 627

\bibitem[{{Cannon}(1973{\natexlab{b}})}]{Can73a}
{Cannon}, C.~J. 1973{\natexlab{b}}, \apj, 185, 621

\bibitem[{{Carciofi} \& {Bjorkman}(2006)}]{Car06}
{Carciofi}, A.~C. \& {Bjorkman}, J.~E. 2006, \apj, 639, 1081

\bibitem[{{Carciofi} \& {Bjorkman}(2008)}]{Car08}
{Carciofi}, A.~C. \& {Bjorkman}, J.~E. 2008, \apj, 684, 1374

\bibitem[{{Dessart} \& {Hillier}(2008)}]{Des08a}
{Dessart}, L. \& {Hillier}, D.~J. 2008, \mnras, 383, 57

\bibitem[{{Dessart} \& {Hillier}(2010)}]{Des10}
{Dessart}, L. \& {Hillier}, D.~J. 2010, \mnras, 405, 2141

\bibitem[{{Dessart} {et~al.}(2012){Dessart}, {Hillier}, {Li}, \&
  {Woosley}}]{Des12}
{Dessart}, L., {Hillier}, D.~J., {Li}, C., \& {Woosley}, S. 2012, \mnras, 424,
  2139

\bibitem[{{Dessart} {et~al.}(2015){Dessart}, {Hillier}, {Woosley}, {Livne},
  {Waldman}, {Yoon}, \& {Langer}}]{Des15}
{Dessart}, L., {Hillier}, D.~J., {Woosley}, S., {et~al.} 2015, \mnras, 453,
  2189

\bibitem[{{Dessart} {et~al.}(2016){Dessart}, {Hillier}, {Woosley}, {Livne},
  {Waldman}, {Yoon}, \& {Langer}}]{Des16}
{Dessart}, L., {Hillier}, D.~J., {Woosley}, S., {et~al.} 2016, \mnras, 458,
  1618

\bibitem[{{Ergon} {et~al.}(2015){Ergon}, {Jerkstrand}, {Sollerman},
  {Elias-Rosa}, {Fransson}, {Fraser}, {Pastorello}, {Kotak}, {Taubenberger},
  {Tomasella}, {Valenti}, {Benetti}, {Helou}, {Kasliwal}, {Maund}, {Smartt}, \&
  {Spyromilio}}]{Erg15}
{Ergon}, M., {Jerkstrand}, A., {Sollerman}, J., {et~al.} 2015, \aap, 580, A142

\bibitem[{{Ergon} {et~al.}(2014){Ergon}, {Sollerman}, {Fraser}, {Pastorello},
  {Taubenberger}, {Elias-Rosa}, {Bersten}, {Jerkstrand}, {Benetti},
  {Botticella}, {Fransson}, {Harutyunyan}, {Kotak}, {Smartt}, {Valenti},
  {Bufano}, {Cappellaro}, {Fiaschi}, {Howell}, {Kankare}, {Magill}, {Mattila},
  {Maund}, {Naves}, {Ochner}, {Ruiz}, {Smith}, {Tomasella}, \&
  {Turatto}}]{Erg14}
{Ergon}, M., {Sollerman}, J., {Fraser}, M., {et~al.} 2014, \aap, 562, A17

\bibitem[{{Esposito} \& {House}(1978)}]{Esp78}
{Esposito}, L.~W. \& {House}, L.~L. 1978, \apj, 219, 1058

\bibitem[{{Falk} \& {Arnett}(1977)}]{Fal77}
{Falk}, S.~W. \& {Arnett}, W.~D. 1977, \apjs, 33, 515

\bibitem[{{Hauschildt} \& {Baron}(1999)}]{Hau99}
{Hauschildt}, P.~H. \& {Baron}, E. 1999, Journal of Computational and Applied
  Mathematics, 109, 41

\bibitem[{{Hillier} \& {Dessart}(2012)}]{Hil12}
{Hillier}, D.~J. \& {Dessart}, L. 2012, \mnras, 424, 252

\bibitem[{{Hillier} \& {Miller}(1998)}]{Hil98}
{Hillier}, D.~J. \& {Miller}, D.~L. 1998, \apj, 496, 407

\bibitem[{{Hubeny} \& {Mihalas}(2014)}]{Hub14}
{Hubeny}, I. \& {Mihalas}, D. 2014, {Theory of Stellar Atmospheres}

\bibitem[{{Jerkstrand} {et~al.}(2015){Jerkstrand}, {Ergon}, {Smartt},
  {Fransson}, {Sollerman}, {Taubenberger}, {Bersten}, \& {Spyromilio}}]{Jer15}
{Jerkstrand}, A., {Ergon}, M., {Smartt}, S.~J., {et~al.} 2015, \aap, 573, A12

\bibitem[{{Jerkstrand} {et~al.}(2011){Jerkstrand}, {Fransson}, \&
  {Kozma}}]{Jer11}
{Jerkstrand}, A., {Fransson}, C., \& {Kozma}, C. 2011, \aap, 530, A45

\bibitem[{{Jerkstrand} {et~al.}(2012){Jerkstrand}, {Fransson}, {Maguire},
  {Smartt}, {Ergon}, \& {Spyromilio}}]{Jer12}
{Jerkstrand}, A., {Fransson}, C., {Maguire}, K., {et~al.} 2012, \aap, 546, A28

\bibitem[{{Kasen} {et~al.}(2006){Kasen}, {Thomas}, \& {Nugent}}]{Kas06}
{Kasen}, D., {Thomas}, R.~C., \& {Nugent}, P. 2006, \apj, 651, 366

\bibitem[{{Kerzendorf} \& {Sim}(2014)}]{Ker14}
{Kerzendorf}, W.~E. \& {Sim}, S.~A. 2014, \mnras, 440, 387

\bibitem[{{Kozma} \& {Fransson}(1992)}]{Koz92}
{Kozma}, C. \& {Fransson}, C. 1992, \apj, 390, 602

\bibitem[{{Kozma} \& {Fransson}(1998)}]{Koz98a}
{Kozma}, C. \& {Fransson}, C. 1998, \apj, 496, 946

\bibitem[{{Kromer} \& {Sim}(2009)}]{Kro09}
{Kromer}, M. \& {Sim}, S.~A. 2009, \mnras, 398, 1809

\bibitem[{{Long} \& {Knigge}(2002)}]{Lon02}
{Long}, K.~S. \& {Knigge}, C. 2002, \apj, 579, 725

\bibitem[{{Lucy}(1991)}]{Luc91}
{Lucy}, L.~B. 1991, \apj, 383, 308

\bibitem[{{Lucy}(1999)}]{Luc99}
{Lucy}, L.~B. 1999, \aap, 345, 211

\bibitem[{{Lucy}(2002)}]{Luc02}
{Lucy}, L.~B. 2002, \aap, 384, 725

\bibitem[{{Lucy}(2003)}]{Luc03}
{Lucy}, L.~B. 2003, \aap, 403, 261

\bibitem[{{Lucy}(2005)}]{Luc05}
{Lucy}, L.~B. 2005, \aap, 429, 19

\bibitem[{{Mazzali}(2000)}]{Maz00}
{Mazzali}, P.~A. 2000, \aap, 363, 705

\bibitem[{{Mazzali} \& {Lucy}(1993)}]{Maz93}
{Mazzali}, P.~A. \& {Lucy}, L.~B. 1993, \aap, 279, 447

\bibitem[{{Olson} {et~al.}(1986){Olson}, {Auer}, \& {Buchler}}]{Ols86}
{Olson}, G.~L., {Auer}, L.~H., \& {Buchler}, J.~R. 1986, \jqsrt, 35, 431

\bibitem[{{Paxton} {et~al.}(2011){Paxton}, {Bildsten}, {Dotter}, {Herwig},
  {Lesaffre}, \& {Timmes}}]{Pax11}
{Paxton}, B., {Bildsten}, L., {Dotter}, A., {et~al.} 2011, \apjs, 192, 3

\bibitem[{{Paxton} {et~al.}(2013){Paxton}, {Cantiello}, {Arras}, {Bildsten},
  {Brown}, {Dotter}, {Mankovich}, {Montgomery}, {Stello}, {Timmes}, \&
  {Townsend}}]{Pax13}
{Paxton}, B., {Cantiello}, M., {Arras}, P., {et~al.} 2013, \apjs, 208, 4

\bibitem[{{Ross}(2007)}]{Ros07}
{Ross}, S. 2007, {Introduction to Probability Models}

\bibitem[{{Scharmer}(1984)}]{Sch84}
{Scharmer}, G.~B. 1984, {Accurate solutions to non-LTE problems using
  approximate lambda operators}, ed. W.~{Kalkofen}, 173--210

\bibitem[{{Shull} \& {van Steenberg}(1982)}]{Shu82}
{Shull}, J.~M. \& {van Steenberg}, M. 1982, \apjs, 48, 95

\bibitem[{{Sim}(2007)}]{Sim07}
{Sim}, S.~A. 2007, \mnras, 375, 154

\bibitem[{{Sim} {et~al.}(2010){Sim}, {Miller}, {Long}, {Turner}, \&
  {Reeves}}]{Sim10}
{Sim}, S.~A., {Miller}, L., {Long}, K.~S., {Turner}, T.~J., \& {Reeves}, J.~N.
  2010, \mnras, 404, 1369

\bibitem[{{Sobolev}(1957)}]{Sob57}
{Sobolev}, V.~V. 1957, \sovast, 1, 678

\bibitem[{{Tanaka} {et~al.}(2007){Tanaka}, {Maeda}, {Mazzali}, \&
  {Nomoto}}]{Tan07}
{Tanaka}, M., {Maeda}, K., {Mazzali}, P.~A., \& {Nomoto}, K. 2007, \apjl, 668,
  L19

\bibitem[{{Verner} {et~al.}(1996){Verner}, {Ferland}, {Korista}, \&
  {Yakovlev}}]{Ver96}
{Verner}, D.~A., {Ferland}, G.~J., {Korista}, K.~T., \& {Yakovlev}, D.~G. 1996,
  \apj, 465, 487

\bibitem[{{Verner} \& {Yakovlev}(1995)}]{Ver95}
{Verner}, D.~A. \& {Yakovlev}, D.~G. 1995, \aaps, 109, 125

\bibitem[{{Werner} \& {Husfeld}(1985)}]{Wer85}
{Werner}, K. \& {Husfeld}, D. 1985, \aap, 148, 417

\bibitem[{{Woosley} {et~al.}(1994){Woosley}, {Eastman}, {Weaver}, \&
  {Pinto}}]{Woo94}
{Woosley}, S.~E., {Eastman}, R.~G., {Weaver}, T.~A., \& {Pinto}, P.~A. 1994,
  \apj, 429, 300

\bibitem[{{Woosley} \& {Heger}(2007)}]{Woo07}
{Woosley}, S.~E. \& {Heger}, A. 2007, \physrep, 442, 269

\end{thebibliography}

\label{lastpage}

\end{document}